%% file: main.tex
\documentclass[sigconf]{acmart} 
\usepackage{subfig}

\AtBeginDocument{%
  }

\acmDOI{XXXXXXX.XXXXXXX}

\begin{document}

% \title{Enhancing Passthrough Comfort with Reprojection}
% \title{Enhancing Passthrough Comfort with Geometry-Aware Reprojection}
% \title{Improving Visually-Induced Discomfort with Geometry-Aware Passthrough}
% \title{Mitigating Visually-Induced Discomfort with Geometry-Aware Passthrough}
% \title{Mind thSe GAP: Geometry-Aware Passthrough Improves Comfort}
% \title{Mind the GAP: Geometry-Aware Passthrough Mitigates Visually-Induced Discomfort}
\title{Geometry Aware Passthrough Mitigates Cybersickness}

\author{Trishia El Chemaly}
%%\authornote{Both authors contributed equally to this research.}
\email{tchemaly@stanford.edu}
\orcid{0000-0002-4234-3082}
%%\author{G.K.M. Tobin}
%%\authornotemark[1]
%%\email{webmaster@marysville-ohio.com}
\affiliation{%
  \institution{Stanford University}
  \city{Stanford}
  \state{CA}
  \country{USA}
}

\author{Mohit Goyal}
\email{mohitgl@google.com}
\affiliation{%
  \institution{Google}
  \city{Mountain View}
  \state{CA}
  \country{USA}
}

\author{Tinglin Duan}
\email{tduan@google.com}
\affiliation{%
  \institution{Google}
  \city{Mountain View}
  \state{CA}
  \country{USA}
}

\author{Vrushank Phadnis}
\email{vrushank@google.com}
\affiliation{%
  \institution{Google}
  \city{Mountain View}
  \state{CA}
  \country{USA}
}
\author{Sakar Khattar}
\email{sakark@google.com}
\affiliation{%
  \institution{Google}
  \city{Mountain View}
  \state{CA}
  \country{USA}
}
\author{Bjorn Vlaskamp}
\email{bjornvlaskamp@google.com}
\affiliation{%
  \institution{Google}
  \city{Seattle}
  \state{WA}
  \country{USA}
}
\author{Achin Kulshrestha}
\email{kulac@google.com}
\affiliation{%
  \institution{Google}
  \city{Toronto}
  \state{ON}
  \country{Canada}
}
\author{Eric Lee Turner}
\email{elturner@google.com}
\affiliation{%
  \institution{Google}
  \city{Cambridge}
  \state{MA}
  \country{USA}
}
\author{Aveek Purohit}
\email{aveek@google.com}
\affiliation{%
  \institution{Google}
  \city{Mountain View}
  \state{CA}
  \country{USA}
}
\author{Gregory Neiswander}
\email{neiswander@google.com}
\affiliation{%
  \institution{Google}
  \city{Mountain View}
  \state{CA}
  \country{USA}
}
\author{Konstantine Tsotsos}
\email{ktsotsos@google.com}
\affiliation{%
  \institution{Google}
  \city{Toronto}
  \state{ON}
  \country{Canada}
}

\renewcommand{\shortauthors}{El Chemaly et al.}

\newcommand{\tbf}[1]{\textbf{#1}}
\newcommand{\directpassthrough}{direct }
\newcommand{\Directpassthrough}{Direct }
\newcommand{\DirectPassthrough}{Direct }

\newcommand{\depthpassthrough}{geometry aware }
\newcommand{\Depthpassthrough}{Geometry aware }
\newcommand{\DepthPassthrough}{Geometry Aware }
\newcommand{\DepthPassthroughAbb}{GAP}

\newcommand{\GAP}{GAP }
\newcommand{\DP}{DP }
\newcommand{\edits}[1]{\textcolor{black}{#1}}
\begin{abstract}
%   Virtual Reality (VR) head-mounted displays (HMDs) provide immersive experiences while limiting the user’s awareness of their physical surroundings. Video see-through (VST) HMDs address this by using outward-facing cameras to reconstruct the user's environment, creating an Augmented Reality (AR) experience. However, relying on video capture from HMD cameras for perception raises concerns about visual discomfort and cybersickness. Since the cameras are positioned outwards and are not located at the eye position, VST HMDs rely on complex image reprojection techniques to create a natural view from the user’s perspective. We first show that direct passthrough (i.e displaying the raw camera feed) leads to exaggerated movements and inaccurate object distances due to inaccurate depth information. Instead, estimating geometry and performing depth-based reprojection can help address these issues but may introduce additional latency and warping artifacts. Using fundamental principles, we discuss a structured approach to designing depth-based passthrough algorithms and introduce metrics to capture warping and perceived geometrical errors. We also design and conduct a user study across 24 participants to compare direct passthrough and depth-based passthrough. Our results demonstrate reduced nausea and disorientation symptoms with depth-based passthrough and uncover several potential avenues to further mitigate visually-induced discomfort.
Virtual Reality headsets isolate users from the real-world by restricting their perception to the virtual-world. Video See-Through
(VST) headsets address this by utilizing world-facing cameras to create Augmented Reality experiences. However, directly displaying
camera feeds causes visual discomfort and cybersickness due to the inaccurate perception of scale and exaggerated motion parallax.
This paper demonstrates the potential of geometry aware passthrough systems in mitigating cybersickness through accurate depth
perception. We first present a methodology to benchmark and compare passthrough algorithms. Furthermore, we design a protocol to
quantitatively measure cybersickness experienced by users in VST headsets. Using this protocol, we conduct a user study to compare
direct passthrough and geometry aware passthrough systems. To the best of our knowledge, our study is the first one to reveal significantly reduced nausea, disorientation, and total scores of cybersickness with geometry aware passthrough. It also
uncovers several potential avenues to further mitigate visually-induced discomfort.
% Virtual Reality headsets isolate users from the real-world by restricting their perception to the virtual-world. Video See-Through (VST) headsets address this by utilizing world-facing cameras to create Augmented Reality experiences. 
% However, directly displaying camera feeds causes visual discomfort and cybersickness due to the inaccurate perception of scale and exaggerated motion parallax. 
% This paper demonstrates the potential of geometry aware passthrough systems in mitigating cybersickness through accurate depth perception. We first present a methodology to benchmark and compare passthrough algorithms. Furthermore, we design a protocol to quantitatively measure cybersickness experienced by users in VST headsets. Using this protocol, we conduct a user study to compare direct passthrough and geometry aware passthrough systems. 
% Our study revealed significantly reduced nausea, disorientation, and total scores of cybersickness with geometry aware passthrough ($p$<0.05). It also uncovers several potential avenues to further mitigate visually-induced discomfort.
\end{abstract}



\begin{teaserfigure}
\centering
  \includegraphics[width=0.88\textwidth]{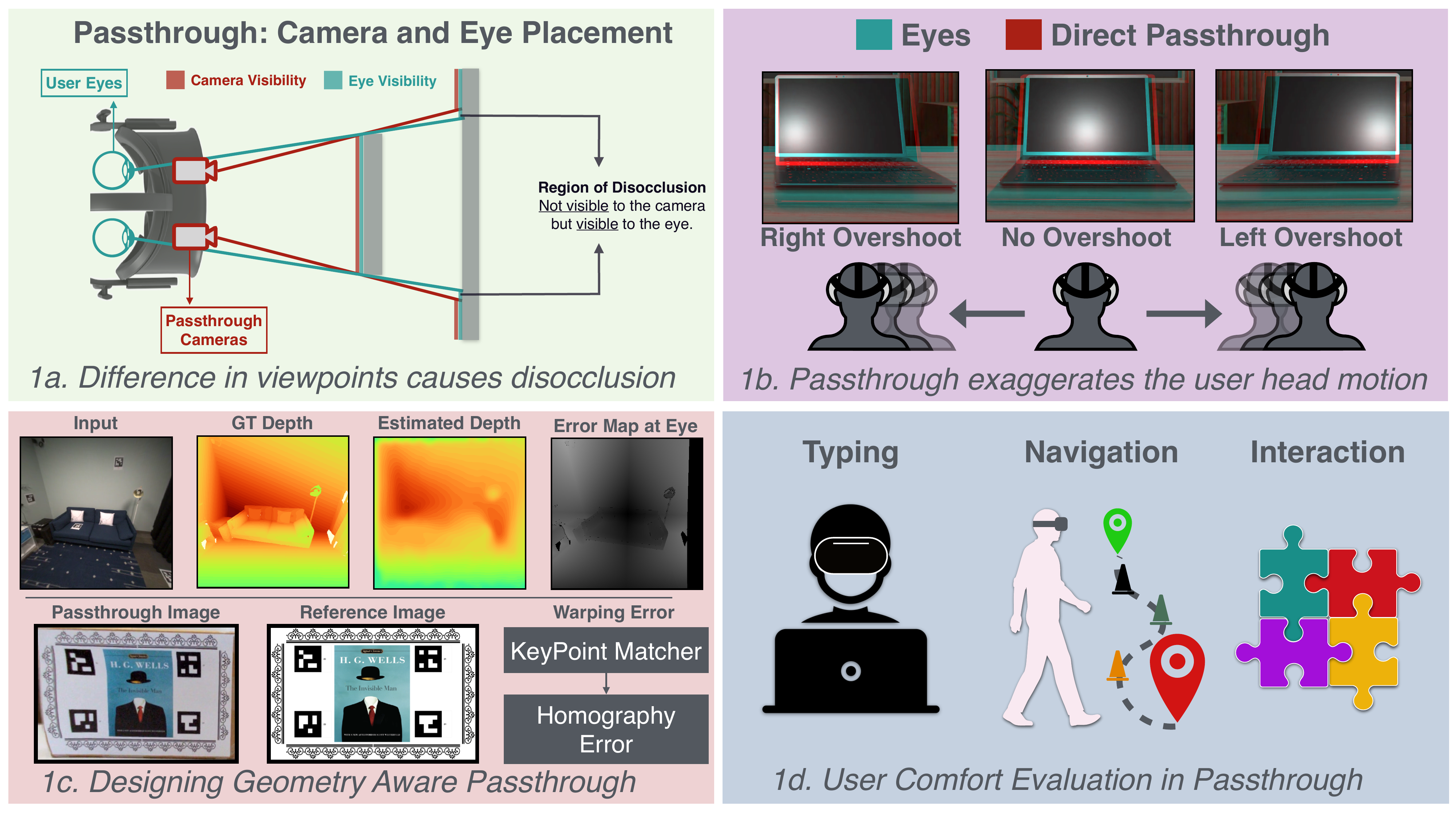}
%   \caption{A typical Video See-Through Headset utilizes high resolution cameras to reveal the physical world to the user. However, due to hardware constraints, these cameras don't capture the same viewpoint as the user's eyes would see without the headset and directly showing camera-feed (direct passthrough) requires solving for disocclusion (1a). Moreover, this difference in viewpoints can also change the perceived objects scale and exaggerate their perceived motion depending on the distance to the cameras (1b). Therefore, geometry aware passthrough algorithms are needed for performing accurate view synthesis that minimizes such errors, and this paper presents metrics to evaluate the perceived object location and any warping induced due to imperfect depth estimation (1c). While these algorithms can improve the accuracy of reprojection, this paper proposes a comprehensive user study based on real-life AR applications and shows the significant impact of this reprojection on cybersickness.}
\caption{Video see-through headsets typically employ high resolution cameras to display the user's physical environment. However, due to inherent hardware limitations, the camera's perspective deviates from the user's natural viewpoint. Hence, directly displaying camera feeds (direct passthrough) can result in visual artifacts such as disocclusion (1a), inaccurate perception of object positions, and exaggerated motion parallax (1b). In this work, we demonstrate that geometry aware passthrough algorithms can circumvent these artifacts, enabling precise view synthesis tailored to the user's eyes. We present metrics for evaluating errors in perceived object location and warping artifacts arising from imperfect depth estimation, fundamental to a seamless passthrough experience (1c). Furthermore, a comprehensive user study is presented to investigate the impact of reprojection algorithms on cybersickness in real-world augmented reality scenarios, highlighting the significance of geometry aware passthrough systems (1d).}
  \Description{Paper Overview: Sub-figure (a) shows a passthrough headset and demonstrates the difference between eye and camera visibility. The difference in viewpoints creates a region of disocclusion where objects are not visible to the camera but visible to the eye. Sub-figure (b) shows three head positions and the corresponding image of a laptop as seen through eyes versus direct passthrough. Passthrough exaggerates motion. Sub-figure (c) demonstrates how we evaluate passthrough algorithms with different metrics. It shows the input image, ground truth depth, estimated depth, and error maps. These are used to calculate object location errors. It also shows a reference and passthrough image. These images are used to calculate warping errors. Sub-figure (d) represents our user study for evaluating cybersickness in VST. It visualizes the three tasks of typing (a person wearing a headset types on a laptop), navigation (a person wearing a headset navigates waypoints and collects cones), and interaction (puzzle). }
  \label{fig:teaser}
\end{teaserfigure}


\maketitle

\input{Sections/Introduction}

\input{Sections/Related-Work}
\input{Sections/Methodology}
\input{Sections/User-Study}
\input{Sections/Results}
\input{Sections/Discussion}
\input{Sections/Conclusion}

\begin{acks}
To Abhishek Kar for his guidance during ideation of this work.
\end{acks}

% \begin{acks}
% \end{acks}

\bibliographystyle{ACM-Reference-Format}
\bibliography{sample-base}

\appendix

\end{document}

%% file: Sections/Introduction.tex
\begin{comment}
To-do list
\begin{itemize}
    \item Polish the intro
    \item  Either use whole word or abbreviation (DP and GAP) everywhere to be consistent
    \item  Related work - VST applications and some previous VST works missing, should be more related to the actual work. No mention of GAP (link reprojection part to GAP?).
    \item Trisha and Vrushank - Finalize thematic analysis
    \item Discussion is missing some relation to previous work
    \item Finalize Conclusion
\end{itemize}
\end{comment}

\section{Introduction}

Virtual Reality (VR) head-mounted displays (HMDs) utilize high resolution near-eye displays and multi-sensory input systems to fully immerse the user in a virtual environment. 
The high level of immersion isolates the user from their physical surroundings. To allow seamless integration between the real and virtual worlds, augmented reality (AR) HMDs overlay digital content onto the reprojected physical world surrounding the user. As a result, in recent years, Video See-Through (VST) or passthrough AR devices such as Apple Vision Pro\footnote[1]{\url{https://www.apple.com/apple-vision-pro}} and Meta Quest 3\footnote[2]{\url{https://www.meta.com/quest}} have become increasingly popular. VST HMDs utilize camera-captured imagery offering users the ability to transition between AR and VR modes to deliver immersive experiences across the full reality-virtuality continuum \cite{milgram1995augmented}. In parallel to VST headsets, Optical See-Through (OST) systems, such as the Microsoft HoloLens\footnote[3]{\url{https://www.microsoft.com/hololens}}, have also been developed. These systems enable users to see the real world through a panel with variable transparency that can display virtual content.

With the rise in popularity of VST HMDs, several AR applications have emerged, including education\footnote[4]{\url{https://giantlazer.com}}, interactive real-world gaming \cite{arGaming}, and especially productivity tasks\footnotemark[1] such as typing, attending meetings, media consumption, and interacting with the physical world in indoor scenarios. As the adoption of these devices as \emph{spatial computers} increases, there is a need for research focused on user discomfort and associated cybersickness with VST technology. While motion sickness in VR \cite{laviola2000discussion, blum2010effect, chang2020virtual} has been extensively studied in last few decades, there is limited work dedicated to enhancing comfort and safety with the use of AR devices. Insights from VR research are informative, but the unique nature of VST where users can see and interact with the physical world suggests a need for dedicated investigation to guide the design of VST HMDs that prioritize user comfort. One of our objectives with this work is to quantitatively measure user discomfort with passthrough systems and provide insights into mitigating user experienced cybersickness.

In VST headsets, due to hardware and physical constraints, cameras are generally positioned on the outer surface of the headset, away from the user's eye positions (see Figure \ref{fig:teaser}a). \emph{Direct passthrough}, i.e. delivering the raw camera feed to the display, exaggerates distances and consequently the motion of objects (see Figure \ref{fig:teaser}b). According to the sensory-conflict theory, the most accepted theory of motion sickness, this mismatch between visual and inertial cues can potentially cause discomfort \cite{oman1990motion}. To mitigate this mismatch, past research focuses on utilizing depth information to reproject camera image feeds into the eye views \cite{chaurasia2020passthroughplus, Lei2022Neuralpassthrough}. We introduce the term \emph{\DepthPassthrough Passthrough} (GAP) to describe these depth-based reprojection algorithms. While previous work assumes that \GAP reduces discomfort compared to direct passthrough, to the best of our knowledge, no empirical studies have directly investigated this assumption. In addition, depth discontinuities and disocclusions can lead to warping artifacts in \GAP algorithms, which could further contribute to the overall cybersickness experienced by users. To address these issues, we propose technical metrics (see Figure \ref{fig:teaser}c) that can estimate the geometrical inaccuracies percevied on the VST displays along with the warping artifacts that potentially modify the shape of objects.

\textbf{Contributions.} In this paper, we first describe the technical differences between \directpassthrough and \depthpassthrough passthrough systems and propose metrics governing the trade-off between enhanced geometrical accuracy and the mitigation of perceived warping artifacts. We demonstrate how these metrics can be used to unveil impact of geometry estimation on the geometry perceived by the user and quantify the introduced distortions in VST images.
% We first provide a structured approach to designing depth-based passthrough algorithms that account for both geometrical accuracy and perceived distortions. 
% Specifically, we introduce two key metrics for evaluating view synthesis in passthrough: (i) spatial reprojection error and (ii) warping error. We utilize these metrics to refine our depth-based passthrough algorithm.
We then introduce a comprehensive protocol (see Figure \ref{fig:teaser}d) focused on key VST use cases to holistically assess visually-induced discomfort and cybersickness in VST HMDs. We apply this protocol to compare our refined \depthpassthrough algorithm to \directpassthrough passthrough. The study encompasses three activities in VST: (i) typing on physical keyboards, (ii) navigation around obstacles, and (iii) near-field object interaction/assembly. Through the proposed study, we first show that VST systems induce several specific symptoms otherwise not experienced without the headset while performing the same tasks. We further demonstrate that \depthpassthrough passthrough significantly reduces nausea, disorientation, and total scores of cybersickness compared to direct passthrough. \edits{By situating our investigation within the context of user comfort, we emphasize the practical implications of GAP for enhancing the human factors and overall user experience of VST HMDs.} We hope our findings will inspire future research focused on designing VST systems with enhanced user comfort. 

Our main contributions in this work can be summarized as follows:
\newcommand{\SubItem}[1]{
    {\setlength\itemindent{15pt} \item[-] #1}
}
\begin{itemize}
\item We propose metrics for quantifying spatial reprojection errors (percevied position and scale) and warping artifacts (change in object shape) which can be utilized to benchmark and compare different passthrough algorithms.
\item We present a protocol for evaluating and quantifying user comfort in VST HMDs. 
\item We demonstrate that \GAP significantly mitigates cybersickness compared to direct passthrough in a user study with 24 participants.
\end{itemize}

    \begin{figure*}[h!]
 \includegraphics[width=\textwidth]{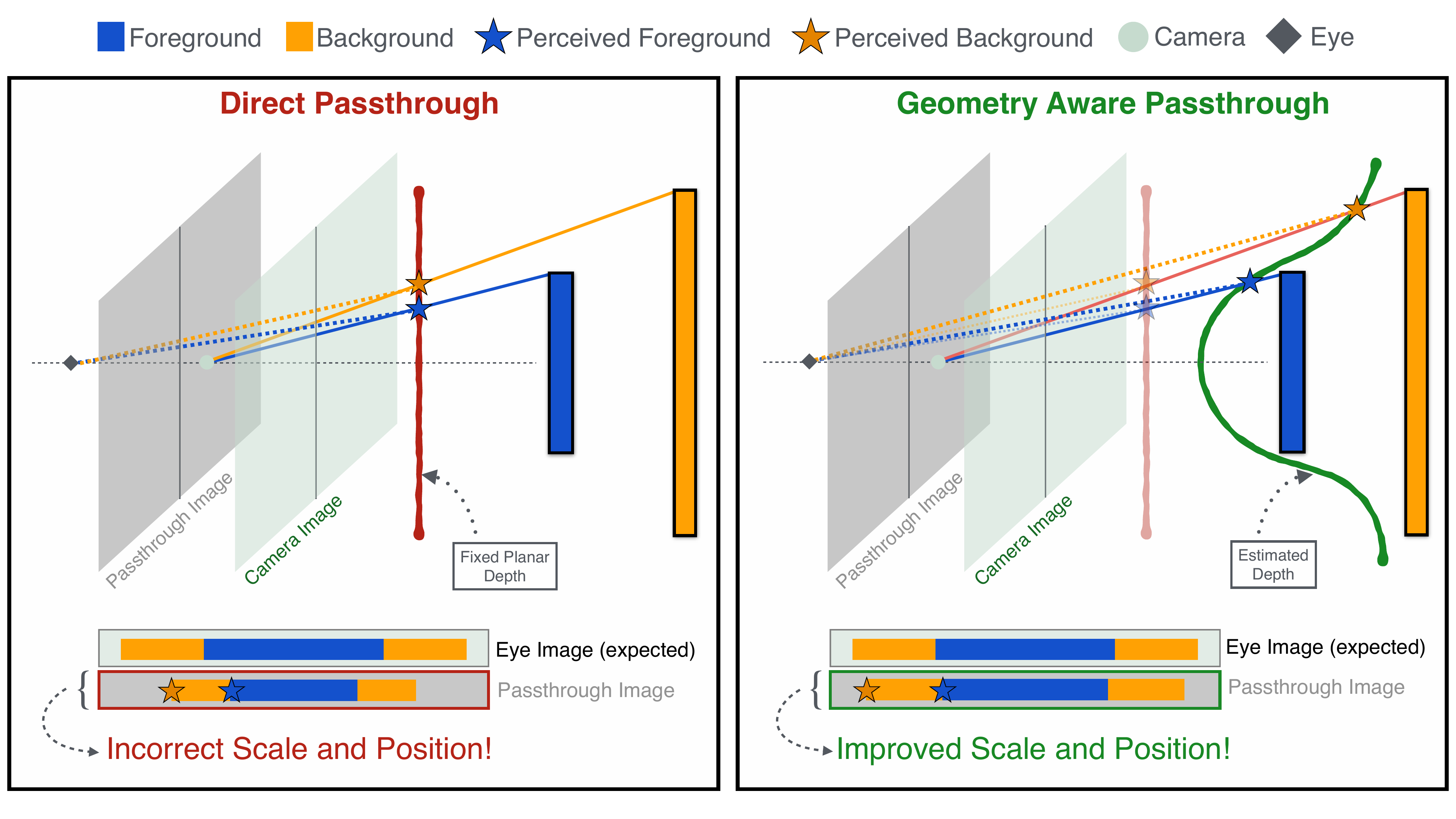}
\caption{\textbf{Comparison between reprojection in \directpassthrough and \depthpassthrough passthrough}. In this figure, we show a 2D illustration to describe the reprojection step where foreground and background are reprojected using different geometries. For direct passthrough, each point in 3D is assumed to belong to a single plane situated at a certain distance. However, that results in reprojecting objects at incorrect pixel locations and at incorrect scale (left). In comparison, \depthpassthrough passthrough uses a geometry estimate which improves the perceived location and the scale (right).}
  \Description{Comparing direct and geometry aware passthrough from first principles. We show a 2D illustration to describe the reprojection where forground (closer objects) and background (far away objects) are reprojected using different geometries. For direct passthrough, each point in 3D is assumed to belong to a single plane situated at a certain distance. However, that results in reprojecting objects at incorrect pixel locations and at incorrect scale. In comparison, geometry aware passthrough uses a geometry estimate which improves the perceived location and the scale. }
  \label{fig:passthroughmodes}
\end{figure*}

%% file: Sections/Related-Work.tex
\section{Related Work}
\subsection{Cybersickness and Comfort in VST}
Literature on VST discomfort builds upon the seminal work on simulator sickness, which informs our current understanding of the psycho-physical causes of cybersickness, such as sensory conflict and postural instability \cite{reason1975motion}. As VR HMDs have become more pervasive over the last few decades, research has validated and expanded our understanding of cybersickness by contextualizing the findings in VR applications \cite{li2022mixed}. LaViola et al.  \cite{laviola2000discussion} studied the visual-vestibular mismatch in VR environments associated with nausea and disorientation and identified vection, the perception of self-motion projected by visual stimuli despite the user being stationary. This effect is particularly pronounced with a wide field of view and rapid changes in the virtual scene, and it may be further accentuated in VST due to increased sensitivity to real-world cues \cite{suwa2022reducing}. Blum et al. investigated out-of-focus blur in VR and found it less problematic, noting lower levels of diplopia (double vision) and higher tolerance for blur \cite{blum2010effect}. They acknowledged that individual differences, such as ocular dominance and susceptibility to motion sickness, influence the accommodation of blur in VR. Additionally, hardware factors like display type, field of view, latency, and graphic realism can contribute to VR sickness \cite{chang2020virtual}. 

Compared to the literature on VR sickness, research on VST-specific discomfort is relatively sparse. Studies have explored various approaches to mitigate simulator sickness in VST systems, including the effects of visual displacement conditions \cite{kim2014effects} and the use of fisheye lenses to expand peripheral vision \cite{orlosky2014fisheye}. Freiwald et al. illustrate the complexities in developing VST technologies by using an offline computing method to create a system to reduce latency \cite{freiwald2018camera}. This approach provided better stabilization, reducing the disconnect introduced by the mismatch between camera and HMD refresh rates. More recently, Li et al \cite{li2022mixed} investigated the effects of mixed-reality tunneling methods on simulator sickness. In our work, we propose a comprehensive evaluation framework that includes both machine-readable metrics and user evaluations (subjective ratings) to address hardware and software aspects of VST. We apply this framework to investigate the effects of geometry aware passthrough on cybersickness, an area that has not yet been examined.
% Compared to the literature on VR sickness, research on VST-specific discomfort is relatively sparse. Hardware and real-time runtime limitations have hindered the full exploration of VST research topics. In a recent example, Freiwald et al. illustrates the complexities in developing VST technologies by using an offline computing method to create a system to reduce latency \cite{freiwald2018camera}. The authors intentionally injected delays in the tracking stream to match the camera system latency. This approach provided better stabilization, reducing the disconnect introduced by the mismatch between camera and HMD refresh rates. 

\subsection{ Novel-View Synthesis and Evaluation.}
    Geometry aware passthrough algorithms fall into the class of novel view synthesis techniques. In contrast to synthesizing views based on fixed camera and eye positions, novel view synthesis aims to solve for a more generalized case i.e. \emph{any} new camera viewpoint given images from a few known viewpoints of the same scene. Early work in this space utilized image-based rendering techniques \cite{SilhouetteIBR, li2023dynibar}], where multiple views were used to construct the scene geometry and then blended together to render novel views. Then, Mildenhall et al. \cite{mildenhall2020nerf} proposed learning a volumetric scene function to model the entire scene which could be queried from novel camera viewpoints. Tretschk et al. \cite{tretschk2020nonrigid} extended this approach to non-rigid scenes, allowing novel view reconstruction over time. Recently, Gaussian splatting \cite{kerbl3Dgaussians} was proposed which aims to model the world with 3D Gaussians and learn these Gaussians to minimize the rendering error on images from known viewpoints. Generally, to evaluate novel view synthesis, novel viewpoints are manually collected using cameras or synthetically rendered. These images then serve as a reference and can be directly compared with the estimated image using metrics like PSNR (Peak Signal-to-Noise Ratio), SSIM (Structural Similarity) and Perceptual Similarity \cite{zhang2018perceptual, kerbl3Dgaussians}. However, these metrics rely on the assumption that reference images are available for evaluation.  \edits{While collecting paired input images of the camera and the eye view is possible \cite{IBRreview}, it is hard to scale for VST headsets and doesn't directly allow assessment from a geometrical standpoint. Instead, we propose metrics that solely utilize depth to compute reprojection errors at the eye, focusing on the geometrical correctness of passthrough systems.}

    \subsection{Reprojection in VST HMDs} For VST HMDs, the passthrough cameras are placed in front of the user's eyes, typically a few centimeters away.
    The scene as viewed by the user's eyes needs to be reconstructed and displayed back to the user through VST displays. This process is generally referred to as \emph{reprojection}, where the camera images are reprojected to the user's eyes. 
    Past work on performing on-device reprojection for passthrough either utilizes classical depth estimation to synthesize eye-views \cite{chaurasia2020passthroughplus} or relies on dedicated GPUs to perform real-time accurate reprojection \cite{novelviewsynthdevice, IBRreview, Lei2022Neuralpassthrough}. In our work, we follow Chaurasia et al. \cite{chaurasia2020passthroughplus}  and implement our geometry aware passthrough pipeline using a low powered, on-device depth estimation algorithm. 
    \edits{While many of these studies discuss several technical aspects that are important for reprojection in headsets, their impact and the significance of reprojection itself on cybersickness remain underinvestigated. With the growing adoption of VST HMDs, we believe that further research into the relationship between reprojection and user comfort is crucial for driving fundamental advancements in VST technology, ultimately enhancing user experience and usability in AR applications.}

%% file: Sections/Methodology.tex
\section{Passthrough Modes and Metric Design}

\begin{figure*}[htbp]
    \includegraphics[width=\textwidth]{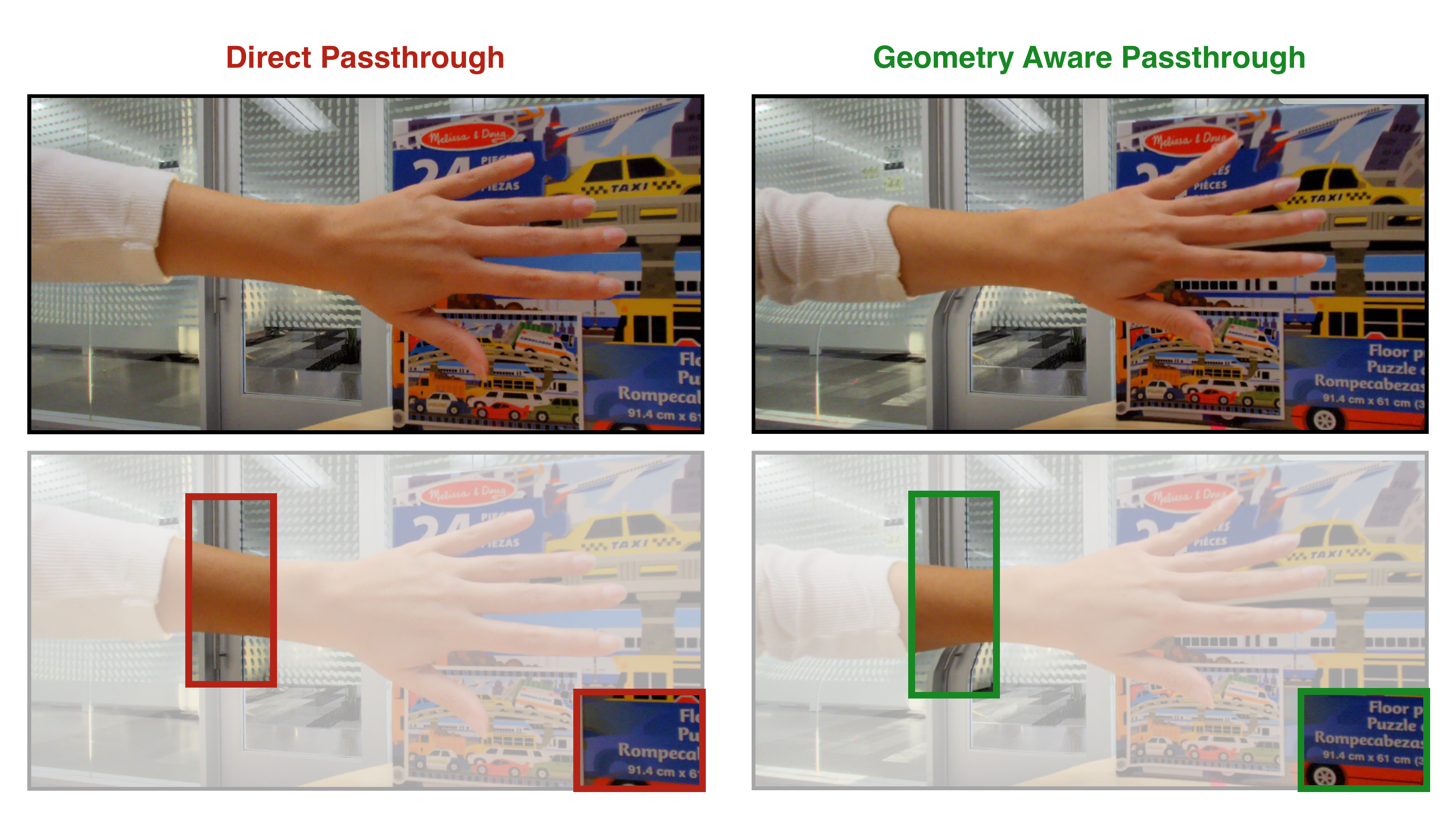}
    \caption{\textbf{\DP versus \GAP output images.} Two images are shown above taken from the headset placed at the same point in the scene. We observe that \DP enlarges all the objects, making the scene look closer to the user. The scale difference can be easily noticed when comparing the text on the bottom right of both images. While \GAP improves the scale, it also results in warping artifacts which can be observed on the edge of the door behind the hand.}
    \Description{We showcase the difference between DP and GAP captures through the headset. In the 4 sub-figures, the capture is that of a hand in front of a puzzle box. The top left capture of DP shows incorrect scale since the puzzle and hand look closer than they actually are. The top right capture of GAP shows warping on the door edge behind the hand. The bottom captures are the same as the top ones with the differences highlighted in squares.}
    \label{fig:artifacts}
\end{figure*}

In this section, we first describe the technical differences between the two passthrough approaches (i) \Directpassthrough Passthrough (DP) and (ii) \DepthPassthrough Passthrough (GAP) that we later evaluate in Section \ref{study}. 
We then provide a technical overview of \GAP and propose metrics for evaluating the geometric accuracy and warping artifacts.
\newline
\textbf{Hardware Details.} The studies were conducted using a mixed-reality headset powered by Android XR\footnote[1]{\url{https://www.android.com/xr/}}.  
We strive to keep all factors other than reprojection such as cameras, image processing, and latency constant across experiments. 
We also utilized the same device to run user studies for fair comparison. 
% We understand that while these factors can have a huge impact on the user comfort, we do not consider these parameters for evaluation in this work and hope that future work can further explore the impact of these factors.

\subsection{Direct Passthrough}

% {\color{red}\textbf{Mohit}}
% \begin{itemize}
%     \item Talk about Direct Passthrough, and problems with exaggerated head motion.
%     \item Emphasize the exaggerated head motion is more prominent on close objects and less on further away objects requiring a depth-based correction.
%     \item Then talk about how this can be seen as a planar re-projection where a fronto-parallel plane is assumed at infinite distance. We instead choose a plane at 1 meters to better reflect the geometry. Later, we will discuss some measure for evaluating geometry, where we will further see the impact of this parameter.
% \end{itemize}

\DP presents unprocessed video feeds directly from external VST HMD cameras, forgoing any form of view correction. This inherent simplicity, however, introduces a critical limitation: visual discrepancies in object positions and scales (see Figure \ref{fig:passthroughmodes}). The disparity between the physical camera locations and the user's eyes manifests, most notably, as an exaggerated perception of motion parallax, particularly pronounced when observing objects in close proximity. 
% The underlying cause of this phenomenon lies in the fundamental difference between the centers of projection for the camera and the human eye.
From a theoretical perspective, \DP can be interpreted as a simplified form of planar reprojection, where the scene is assumed to be geometrically planar (see Figure \ref{fig:passthroughmodes}), with the plane being fronto-parallel and situated infinitely far from the cameras. In our practical implementation, we modify the positioning of the projection plane at a distance of 2 meters from the cameras since most of the objects lie in the 1-2 meters range. This adjustment helps improve the perceived scale of the scene by \emph{better} reflecting the average depth of objects, specifically in context of the activities chosen for cybersickness analysis (see section \ref{subsec:tasks} for more details). Moreover, it can be implemented using a simple homography transformation with minimal computational cost, without doing any geometry based correction fundamental to \DP.
% Direct Passthrough approach directly displays the video captured by external VST HMD cameras without any modifications such as reprojection techniques. This approach does not account for the difference in physical positions of the external cameras and the eyes, and presents the video directly to the user without accounting for the visual errors introduced by camera-eye separation. The differences in center of projections for camera and eye can produce perceived exaggerated movement due to head motion The exaggerated head motion is especially prominent on close objects and less far objects, requiring a depth based correction. This method is analogous to a planar re-projection where a fronto-parallel plane is assumed at infinite distance. We instead choose a plane at 1 meters to better reflect the geometry. 
% Later, we will discuss some measure for evaluating geometry, where we will further see the impact of this parameter.

\begin{figure*}[ht]
  \centering
  \includegraphics[width=\linewidth]{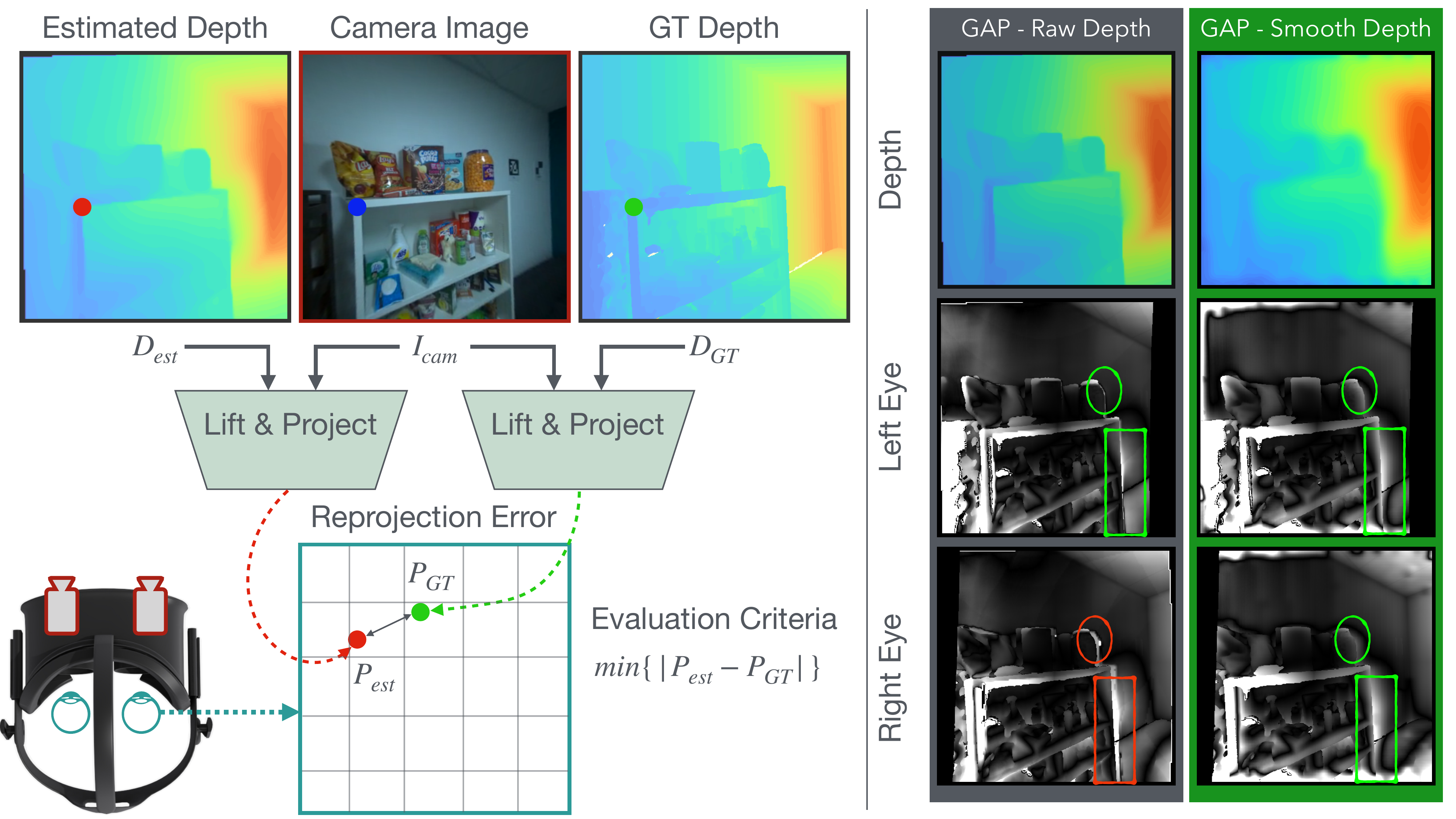}
  \caption{\textbf{Spatial Reprojection Error}. Evaluation of perceived scene geometry with reprojection error. We take the Estimated Depth and GT Depth for the same image, and reproject every pixel to the eye view. The differences in the reprojected locations is used as the evaluation criteria. On the right, we compare the effects of smoothing the depth. Since the estimated depth can have errors, warping and rotating this depth map to the right camera can further exacerbate these errors. Gaussian smoothing of the estimated depth helps reduce these errors on the right eye as shown above.}
  \label{fig:geometrical_metric}
  \Description{Geometrical Assessment Measure: On the left, we visualize how the spatial reprojection error is calculated. Heat maps of the estimated and ground truth depth are shown. We also visualize the distance between the projected ground truth and estimated points on a grid. This distance is the reprojection error. On the right, we visualize heat maps of depth with raw and smooth GAP.}
\end{figure*}

\subsection{\DepthPassthrough Passthrough}
In contrast to \DP, \GAP employs novel view synthesis techniques or reprojection to align external camera images with the user's viewpoint, thereby mitigating the inherent disparity between camera and eye positions. 
While this reprojection can potentially enhance visual fidelity, it requires more computational resources, which can introduce additional latency into the VST system. 
Furthermore, even under ideal conditions with perfect geometric estimations, the reprojection process may result in gaps or holes in the rendered imagery due to depth disocclusion. Disocclusions occur when objects closer to the camera obscure portions of the scene that would be visible from the user’s perspective. 
\GAP is susceptible to two primary categories of artifacts. First, geometric artifacts can manifest due to inaccurate depth perception, leading to incorrect positioning and scale of objects. 
Second, warping artifacts can emerge as a consequence of filling the aforementioned gaps or holes by stretching the foreground or background at depth discontinuities (see Figure \ref{fig:artifacts}). 
% The severity of these warping artifacts is contingent upon the quality and accuracy of the depth estimation algorithms utilized.
In our specific implementation, we leverage a stereo depth estimation algorithm to obtain the depth information for the left camera. We then generate the corresponding depth maps for the right camera view using known camera calibration parameters. It is important to note that this choice of depth estimation method is not fundamental to GAP; it can be replaced with any other suitable depth estimation technique without compromising the overall methodology. We follow Chaurasia et al. \cite{chaurasia2020passthroughplus} as our approach is similar to their passthrough implementaiton ---  \emph{Passthrough+}.
% To address this, warping techniques are needed to fill these missing areas, but this can introduce additional visual artifacts. The potential errors in the eye space resulting from this approach are primarily two-dimensional in nature: (i) Shifts or scaling artifacts can distort the original geometry of the scene, and (ii) warping artifacts can arise depending on the accuracy of the depth estimation algorithms.

\input{geometrical_metrics}

\subsubsection{Geometry Evaluation}
\edits{Geometrical correctness of the rendered scene is fundamental to any AR/VR system \cite{de2024visual, deptheval2023, objectdepth22} that relies on the head tracking quality, depth estimation, device calibration and the overall rendering pipeline. Particularly for passthrough,} the main advantage of using geometrical knowledge of the scene before applying reprojection is to \edits{allow} for correctly projecting each point in the world exactly where the eye expects to see it on the display. However, geometry estimation is not perfectly accurate, and the estimate needs to be reused for multiple future images (to perform reprojection in real time), which further impacts its accuracy. Therefore it is crucial to evaluate the impact of errors in depth estimation on the display as seen by the user. We demonstrate that these errors cannot be trivially obtained by just examining geometry errors in 3D space, such as mean absolute error in depth estimated per pixel. To this end, we obtain ground-truth depth of several indoor scenes using 3D laser scanners and then calibrate our headset in those scenes using ArUco \cite{aruco} markers with a bundle adjustment algorithm to collect aligned benchmarking data. We collected 9 datasets in total, each containing 3000 images. This provides paired frames $D_{est}, D_{GT}$ for every input camera image $I_{cam}$ (our metric does not rely on $I_{cam}$ and we use it for visualization purposes only). Then, each pixel in $I_{cam}$ is reprojected into the eye view using the reprojection operator which uses the depth and the camera intrinsics for the input camera, followed by projection into the eye. We define the resulting pixel projections as $P_{est}$ and $P_{GT}$ when using $D_{est}$ and $D_{GT}$, respectively. Finally, the error is computed as the L-1 norm of the residual $|P_{est} - P_{GT}|$ (see Figure \ref{fig:geometrical_metric}). We refer to this as the spatial reprojection error, which we compute for both left and right eye. Additionally, we compute the mean absolute error in depth as $|D_{est} - D_{GT}|$ averaged over the image.

\begin{figure}[!h]
    \centering
    \includegraphics[width=\linewidth]{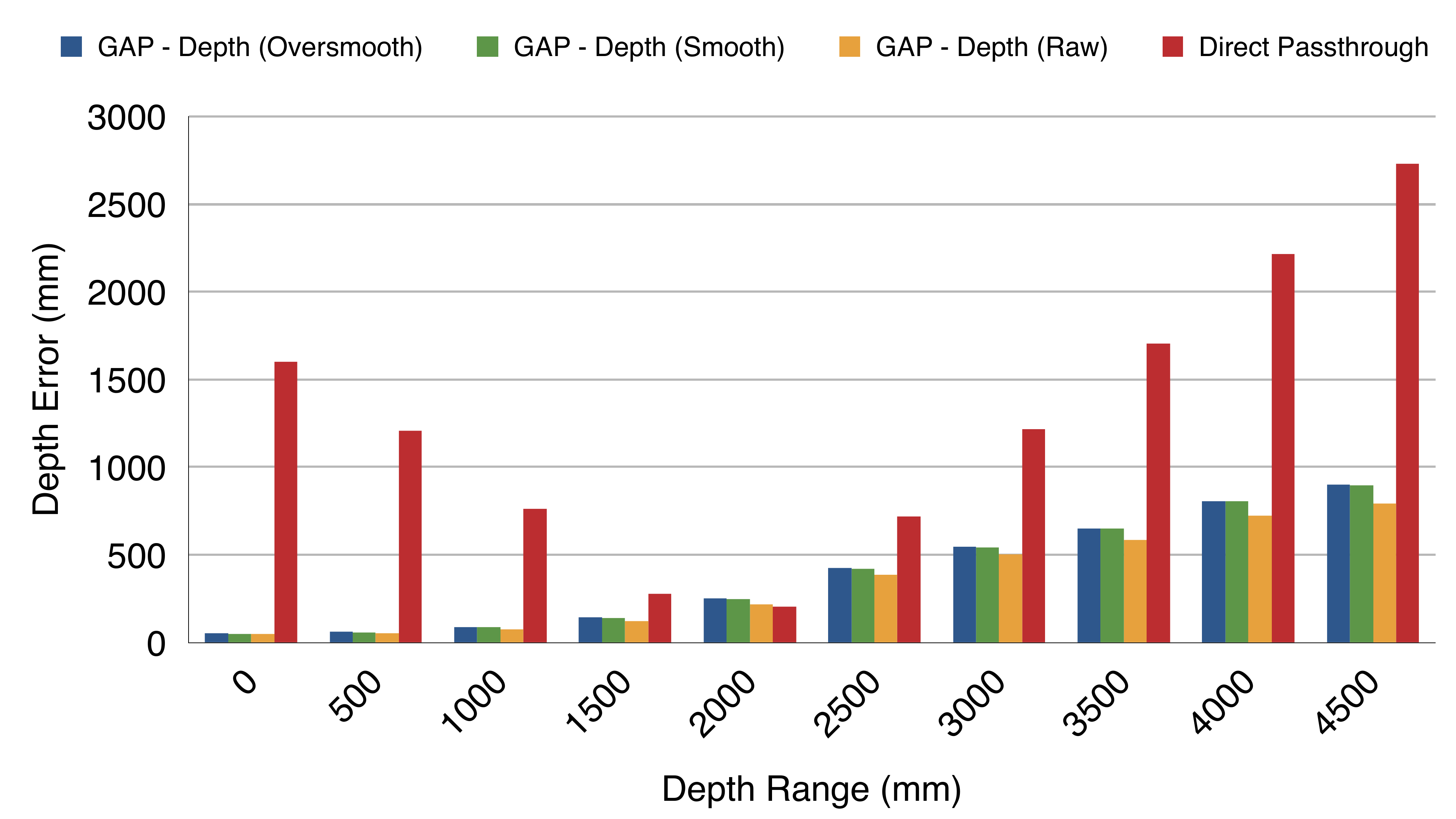}
    \caption{\textbf{Geometrical Errors.} Comparison of the errors in depth estimation for different \GAP variants and DP. We observe that smoothing increases the depth errors. DP on the other hand has the highest error among all approaches achieving minimum depth error at the 2m (2000mm) mark which is expected since we assume all the points are at 2m for DP.}
    \Description{Bar plot of GAP (raw), GAP (smooth), GAP (oversmooth), and DP depth error. The y-axis represents depth error (mm) while the x-axis represents depth range (mm). All GAP methods show an increasing depth error with increasing depth range. Smoothing increases depth errors. DP has the highest error among all approaches. The error with DP decreases until 2000 mm then increases again.}
    \label{fig:depth_errors}
\end{figure}

We compared \DP and \GAP using the spatial reprojection error and the results are summarized in Table \ref{tab:geometric_metrics}. We observed that DP, on average, has a 3.2 pixel error across the 9 datasets we evaluated on. In comparison, \GAP - Depth (Raw) gave a 4x improvement in the spatial reprojection error (see differences in perceived scale in Figure \ref{fig:artifacts}). However, we observed that the error on the right eye is much higher than that on the left eye since the depth is estimated on the left camera and then warped to the right camera. We believe that this depth warping can exacerbate the depth errors on objects, particularly at object boundaries due to depth discontinuities. This is consistent with our observation that \GAP - Depth (Smooth), which uses a Gaussian kernel, reduces the error on the right eye by 25\%. Since smoothing smoothes out changes in depth, we see that errors due to warping get reduced specifically on object edges in the right eye after smoothing (see spatial reprojection errors visualized on the right in Figure \ref{fig:geometrical_metric}). However, when we increased the smoothing of the estimated depth further, we observed a higher spatial reprojection error in both eyes. We emphasize that these insights, derived from spatial reprojection error, cannot be  understood by examining depth error alone in 3D space (as seen in Table \ref{tab:geometric_metrics}). We also observed the error in depth estimation at different scene depths for DP and different variants of GAP in Figure \ref{fig:depth_errors}. We noted that DP achieves  minimal depth error only for points approximately 2 meters in depth, which supports our findings. This further validates the choice of GAP over DP. For the remainder of the paper, we refer to Geometry Aware Passthrough with smooth depth as \GAP since it achieves the best spatial reprojection error for both eyes and Direct Passthrough as \DP.

\begin{figure*}[!h]
  \centering
  \includegraphics[width=\linewidth]{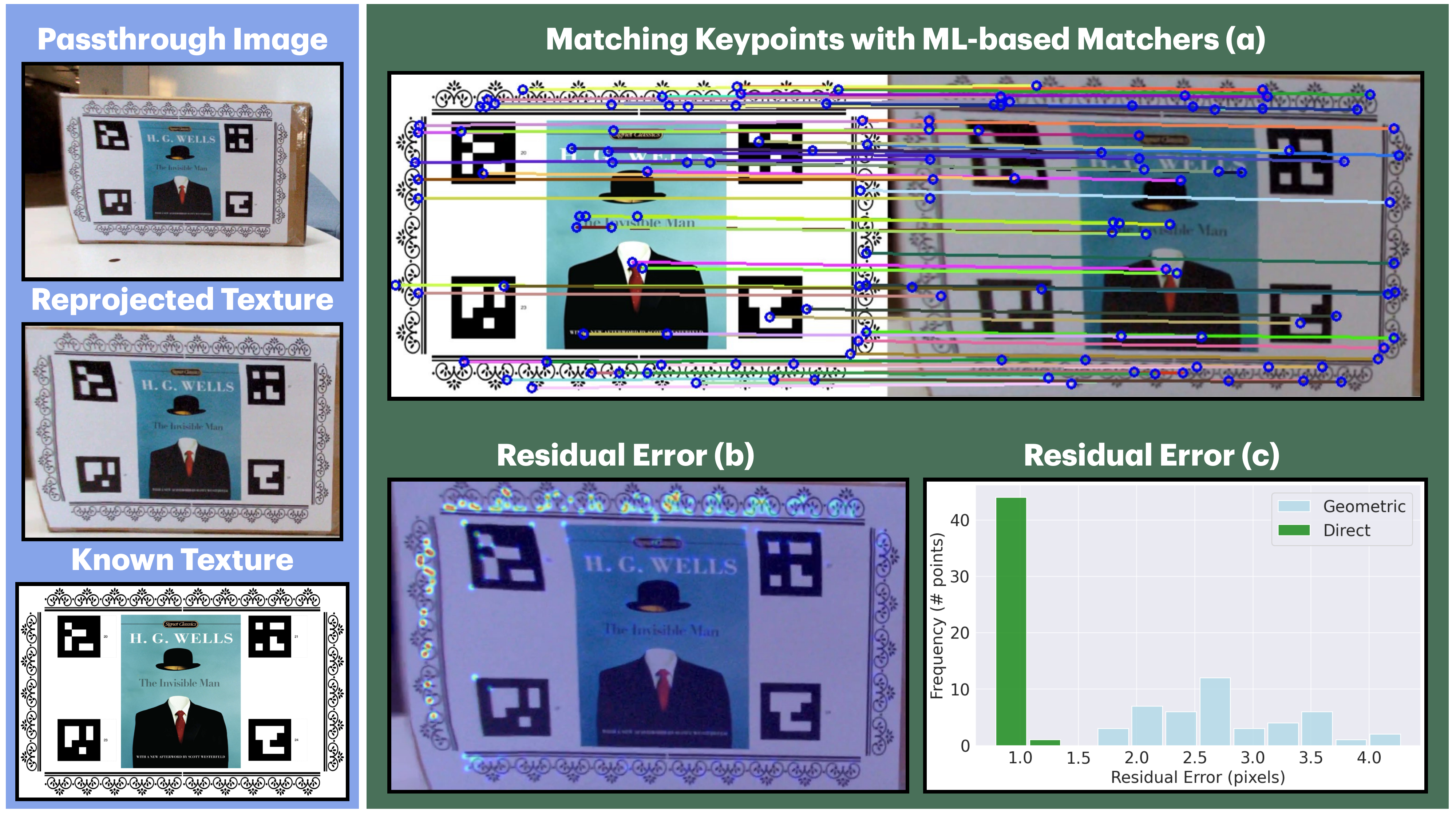}
  \caption{\textbf{Warping Errors.} We consider planar targets with known hand-crafted textures to quantify warping artifacts in the passthrough images. Specifically, we take the passthrough reprojected image and obtain the crop around the reprojected texture of interest using ArUco \cite{aruco} marker detectors (see left tile). Then we obtain the correspondences between the known texture and the reprojected texture using ML-based keypoint matching techniques \cite{sarlin20superglue} (a). We then use a homography solver to find the residual errors which measures the deviation of the reprojected image from a planar surface. This residual error can be plotted on the image (b) indicating pixel locations where the warping is observed. We also observe that \depthpassthrough passthrough has a higher warping error than \directpassthrough passthrough (c).}
  \Description{The figure demonstrates how warping errors are calculated. A reference image as well as passthrough image with its corresponding cropped image are shown. We visualize the keypoints matched between the reference and cropped image. We also show a bar plot of the residual error. The y-axis represents frequency (number of points) and the x-axis represents residual error (pixels). The frequency for DP is high at a residual error of 1 pixel. The frequency for GAP is spread over the residual error range of 1 to 4 pixels.}
  \label{fig:warping_metric}
\end{figure*}

\subsubsection{Warping Evaluation}

While \GAP improves the perceived location of objects in the scene (as shown in Figure \ref{fig:passthroughmodes}), the errors in geometry estimation introduce warping artifacts. This manifests as bending artifacts where rigid objects can appear to deform, like jelly, altering their perceived shape. This issue becomes immediately noticeable to users when using GAP (more in Section 4 and see). Therefore, we propose a simple method to quantify these effects for any black-box system by directly using the synthesized imagery. While spatial reprojection metric uses depth estimates to compute reprojection errors, this metric uses the final image output by the passthrough device to quantify stretching on objects of know shape and texture. Specifically, we choose planar targets with hand crafted textures and compute correspondences between known and reprojected textures. For estimating correspondences, we employ SuperGlue \cite{sarlin20superglue} which uses deep neural networks to extract features and graph-based algorithms to match them across images. Then, we use RANSAC to estimate a homography between these correspondences as they represent a planar target. \edits{The accuracy of this formulation highly depends on the quality of keypoint matching and stability of homography estimation. To ensure high quality matching, we utilize the state of the art ML-based method for detecting keypoints. To stabilize the homography estimation, we adhere to two principles: obtaining a large number of matches and ensuring the matches are of high quality. To achieve this, we used a custom-designed target with substantial texture variation and applied a confidence threshold of 0.3 to obtain at least 100 high-quality matches, enabling stable homography estimation. Additionally, to further reduce noise, we averaged the residuals over 45 frames to calculate the final metric.} We finally report the residual error from the resulting fit output by the RANSAC algorithm (see Figure \ref{fig:warping_metric}).

\input{warping_metrics}

As shown in Table \ref{tab:warping_metrics}, we indeed observe that \GAP introduces significantly more warping compared to \DP. We did not find meaningful reduction in these artifacts on the pasted textures through different depth estimates (raw, smooth and oversmooth). While the proposed metric accurately quantifies bending introduced by GAP, we believe that counteracting and evaluating these issues require \edits{more complex models than just homography on planar targets and} more complex metrics that can capture not only for static images, but also their temporal nature.

\edits{\textbf{Impact of warping artifacts on user experience:} Through discussions with passthrough users, we identified that warping artifacts substantially affect their preferences and overall experience. We note that an important consideration of our approach is applicability to any passthrough system without requiring access to the internal state of the system. The subjective responses from participants, discussed in Subsection \ref{subsec:qualfeedback}, provide insight into the effect of warping artifacts on user experience and comfort.}

% It’s equally important to measure the warping artifacts induced in the Depth Passthrough approach, but directly measuring the depth errors is not enough to capture warping effects. Since generalizing to unknown scenes with unknown geometry is hard, we propose a simplification and estimate warping in the reprojected imagery on objects of known geometry and texture. We specifically choose planar targets with hand crafted textures and compute correspondences between known and reprojected textures. These correspondences are then validated to be in geometrical agreement to a planar surface. The residual errors (measured in pixels), if any, indicate the induced warping.

%% file: geometrical_metrics.tex
\begin{table*}[htbp]
  \caption{Comparison for different reprojection methods. We report reprojection error (in pixels) averaged over 9 datasets that we obtained ground truth 3D recordings for.}
  \label{tab:geometric_metrics}
  \resizebox{0.8\textwidth}{!}{
  \begin{tabular}{rc|c|c|c}
    \toprule
    Passthrough Mode&\multicolumn{2}{c}{Spatial Reprojection Error}&\multicolumn{2}{c}{Depth Error}\\
    \midrule
    & Left Eye (pixels) & Right Eye (pixels) & Left Eye (meters) & Right Eye (meters) \\
    \midrule
    Direct Passthrough &  3.19 $\pm$ 2.56 & 3.15 $\pm$ 2.32 & 2.62 $\pm$ 3.03 & 2.62 $\pm$ 3.00 \\
    GAP - Depth (Raw)                   &  0.68 $\pm$ 0.50 & 1.05 $\pm$ 0.46 & \tbf{2.03} $\pm$ \tbf{3.08} & \tbf{2.03} $\pm$ \tbf{3.08}\\
    GAP - Depth (Smooth)                &  \tbf{0.68} $\pm$ \tbf{0.51} & \tbf{0.75} $\pm$ \tbf{0.53} & 2.05 $\pm$ 3.10 & 2.04 $\pm$ 3.07\\
    GAP - Depth (Oversmooth)            &  0.72 $\pm$ 0.53 & 0.78 $\pm$ 0.53 & 2.05 $\pm$ 3.10 & 2.05 $\pm$ 3.07  \\
  \bottomrule
\end{tabular}
}
\end{table*}

%% file: warping_metrics.tex
\begin{table}[!ht]
  \caption{Comparison for different reprojection methods. We report residual warping error (in pixels) averaged over 5 second videos spanning 45 frames. A higher warping error indicates more warping.}
  \label{tab:warping_metrics}
  \resizebox{\columnwidth}{!}{
  \begin{tabular}{rc|c|c}
    \toprule
    Passthrough Mode&\multicolumn{3}{c}{Warping Error (pixels)}\\
    \midrule
    & Mean & Median & 90th Percentile \\
    \midrule
    DP  & 0.96 $\pm$ 0.08 & 0.95 & 1.73\\
    GAP      & 2.71 $\pm$ 0.95 & 2.57 & 6.90           \\
  \bottomrule
\end{tabular}
}
\end{table}

%% file: Sections/User-Study.tex
\section{User Study} \label{study}

As AR headsets are becoming popular with an increasing number of applications in medicine, education, and gaming, it becomes important to holistically evaluate the discomfort experienced by users in these passthrough systems. To address this, we introduce a protocol specifically tailored to quantify cybersickness in the context of VST HMD use cases. \edits{To our knowledge, no VST studies comprehensively combine user motion, interaction, and sickness metrics, making this protocol an early effort to define reasonable benchmarks.}

\edits{\subsection{Experiment Design Considerations}
{Our key considerations for the experimental setup were reproducibility, repeatability, and real-life relevance while the study design focused on reliably eliciting signals related to user comfort. To achieve this, we began with tasks identified in the literature, tested them in a pilot study, and iteratively refined the task nature and duration based on participant feedback. We detail the specific steps we took to ensure these principles were upheld:
\subsubsection{Standardization of Setup and Metrics}
{We standardized all aspects of the experimental setup to ensure reproducibility and consistency. This included the layout of the room, object placement, and controlled environmental factors such as lighting and headset parameters. We provide this information in our Supplementary material. For cybersickness and discomfort measurements, we employed established tools which are widely validated in prior literature.}
\subsubsection{Pilot Testing and Iterative Refinement}
Pilot studies played a crucial role in refining task design and duration to reliably induce motion sickness while maintaining ecological validity. Tasks were initially drawn from existing research that we summarize in our Supplementary material and adjusted based on observed results and participant feedback. Specific triggers of motion sickness such as head motion, depth perception tasks, and visual-motor coordination were emphasized in the task design. For example, we replaced straight-line walking tasks with multi-directional navigation requiring turns and object interactions to better simulate realistic scenarios that elicit depth perception challenges and frequent head motion.
\subsubsection{Real-Life Relevance}
The protocol was designed to mirror real-world VST use cases, ensuring relevance to everyday applications. Tasks included typing on a physical keyboard for productivity, navigating complex physical environments, and interacting with tangible objects, such as completing assembly tasks. This mix of stationary (near-field interaction) and dynamic (locomotion-based) activities ensured a comprehensive evaluation of user discomfort across different contexts.
}}

Below, we outline the design of our protocol, including task selection and quantification methods. We then apply our protocol to understand the effects of the reprojection algorithms on discomfort and cybersickness when using a VST HMD. Given the subjective nature of cybersickness and discomfort, we employ a within-subject design in our protocol where participants complete three tasks involving head motion, hand-eye coordination, and untethered locomotion. We collect both objective task performance metrics and quantitative and qualitative subjective feedback under three conditions: Natural Vision (NV), \Directpassthrough Passthrough (DP), and \DepthPassthrough Passthrough (GAP).

\begin{figure*}[ht]
    \centering
    \includegraphics[width=\textwidth,trim={0.2cm 10cm 0.2cm 0.2cm},clip]{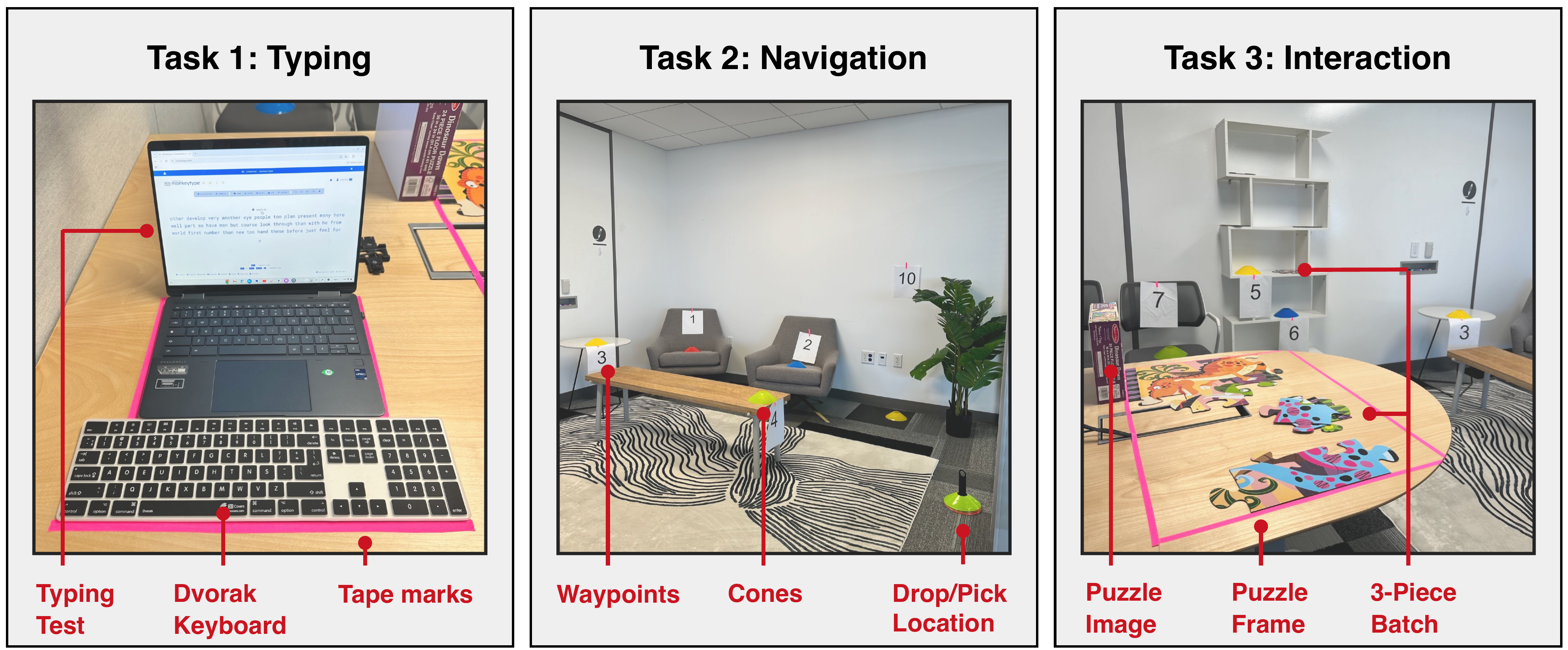}
    \caption{\textbf{User Study Setup.} Pictures of the lab setup for the three tasks completed by the participants while wearing the VST HMD.}
    \Description{The figure shows the user study setup in the lab space. The left sub-figure shows the setup for the typing task: a laptop and Dvorak keyboard. Their location is marked with tape for consistency between participants. The middle sub-figure shows the setup for the navigation task. The cones are numbered and placed around pieces of furniture such as a plant, 2 tables, and 2 chairs. The right sub-figure shows the setup for the interaction task. Puzzle pieces are placed on shelves. The table is marked with tape to indicate the frame in which the completed puzzle should fit.}
    \label{fig:tasks}
\end{figure*}

\subsection{Participants}
For the user study, we recruited 25 participants who had normal or corrected-to-normal vision. The study included a diverse range of ages, genders, levels of VR usage, and job profiles. Table \ref{tab:demographics} gives a full overview of the demographics. Recruitment was done in accordance with the ethics board of our institution. We excluded data from one participant from our analysis since they spent an abnormally long time to complete the protocol. \edits{Since the rest of the participants exhibited no irregular behavior, all their data were retained for analysis including participants who reported higher than average sickness scores.}

\begin{table}[!ht]
  \caption{Participant demographics for the user study, showing diversity across gender, age, normal versus corrected vision, and VR usage.}
  \label{tab:demographics}
  \begin{tabular}{l|l|l}
    \toprule
     \textbf{Variable} & \textbf{Categories} & \textbf{\#Participants}\\
    \midrule
    \textbf{Gender} &Men &14\\
    & Women &11\\
    \midrule
    \textbf{Age Group} &18-24 &1\\
    &25-34 &12\\
    &35-44 &9\\
    &45-54 &3\\
    \midrule
    \textbf{Vision} &Normal &19\\
    &Contact Lenses &6\\
    \midrule
    \textbf{VR Usage} &Never &1\\
    &Once &12\\
    &Once a week &3\\
    &Once a month &8\\
    &At least once a day &1\\
    \bottomrule
  \end{tabular}
\end{table}

% (14 males, 11 females). 1 participant was 18-24 years old, 12 were 25-34 years old, 9 were 35-44 years old, and 3 were 45-54 years old. None of the participants were over 55 years old. 
% Participants had normal or corrected-to-normal vision (19 participants had normal vision and 6 participants wore contact lenses). 1 participant never used VR before, 12 participants used VR once, 3 participants use VR once a week, 8 participants use VR once a month, and 1 participant uses VR at least once a day. 
\subsection{Tasks}\label{subsec:tasks} 

We devised our protocol focusing \emph{exclusively} on passthrough-based real-world interactions and ensured no virtual elements \cite{de2024visual} were visible to participants. Figure \ref{fig:tasks} shows pictures of the lab setup for the three tasks completed by the participants. The tasks were inspired from fundamental real-world AR applications such as working with laptops for productivity, navigation in the physical world, and interaction with real-objects. They emphasized on the user head motion while  necessitating inspection and spatial awareness of the physical world.
% head motion, realistic interaction, and spatial awareness within a physical environment. 
Our design ensured a cumulative duration to be roughly 15 minutes for all the tasks.

\subsubsection{Typing} Typing is a familiar activity that effectively engages both visual and motor components, making it relevant for evaluating fine-selecting physical objects and digital screen usage. \edits{This task was specifically chosen to reflect emerging applications in productivity such as using VST HMDs like the Apple Vision Pro as extensions of traditional work setups where users interact with physical keyboards and screens. It is representative of fine motor interactions requiring frequent gaze shifts between the keyboard and the laptop screen, a common trigger for visual discomfort. } A physical Dvorak keyboard \cite{dvorak1936typewriting} was used to minimize reliance on muscle memory and encourage frequent visual engagement with the keys. A typing assessment was conducted to measure speed, accuracy, and overall proficiency. Participants typed for 6 minutes as quickly and accurately as possible, with the typing text randomized for each session to ensure variability. \edits{To maintain repeatability and reproducibility, the placement of the laptop and keyboard was standardized across all participants, and screen brightness was kept consistent throughout the study.}
\subsubsection{Navigation} The realistic and holistic use of a VST HMD involves navigating physical spaces, avoiding obstacles, and interacting with real-world objects \cite{bailenson2024seeing,de2018augmented,erickson2019cold}. Previous research has also explored the impact of spatial navigation, such as waypoint-following, on sickness in VR  \cite{al2019effect}. We designed a navigation task where participants collected and dropped off 10 numbered cones, one at a time, into a designated drop zone. \edits{To emphasize geometry perception, the task included narrow passages and required multi-directional movements, including turning to locate cones and interacting with objects at varying heights.} Further, the cones were placed at different heights, requiring a range of movements to pick and place. Participants were given 2-3 minutes to familiarize themselves with the cone locations before beginning the trial. During the task, participants were instructed to move naturally but carefully to avoid colliding with any objects in the room.
\subsubsection{Interaction} Li et al. \cite{li2022mixed} introduced a Tangram puzzle task as a representative activity that simulates common assembly tasks requiring both motor and cognitive skills. Inspired by their approach, we adapted this task by using jigsaw puzzles consisting of 24 pieces and measuring 2 x 3 feet. This larger puzzle size was selected to accommodate head motion which is often associated with motion sickness. We chose a 24-piece puzzle to mitigate prolonged VST exposure and limited the time of each condition to a maximum of 20 minutes.

To maintain a consistent level of difficulty across conditions, we utilized three different jigsaw puzzles from the same series by the same manufacturer. To standardize the puzzle completion strategy, we divided the puzzle pieces into 8 batches placed on the same shelves for all trials. Participants were restricted to retrieving and working on only one batch at a time within a rectangular frame marked on the table. This setup allowed us to incorporate locomotion into the task as participants moved between batches. The puzzles were assigned to the conditions randomly but were not repeated.

\subsection{Measures}
\subsubsection{Cybersickness} 
Although multiple VR sickness measurement schemas such as ARSQ, VRSQ, and CSQ-VR have been proposed, the Simulator Sickness Questionnaire (SSQ) remains the most prevalent in the literature and is still widely regarded as the standard for measuring VR sickness \cite{vinkers2024visual, kourtesis2023cybersickness, kim2018virtual, hussain2023augmented, hirzle2021critical}. SSQ is a self-reported checklist consisting of 16 symptoms categorized under four subscales: Nausea, Disorientation, Oculomotor, and Total Severity. 
%To assess the feasibility of including alternatives to the SSQ, we included the digital eye strain (DES) measure to augment the SSQ in our pilot studies \cite{}. However, our initial findings did not find the DES to be as effective as the SSQ. As a result, in our work, we primarily rely on the SSQ.

Participants rated the severity of these symptoms on a 4-point Likert scale (0 = none, 1 = slight, 2 = moderate, 3 = severe). SSQ scores were collected right before and after each condition. This allowed us to isolate the SSQ per condition by evaluating the difference between post and pre task SSQ scores, as evidenced in prior literature \cite{li2022mixed}. 

We also collected discomfort scores by asking participants to rate their discomfort immediately after completing each task while still wearing the headset. At the beginning of the study, participants were informed that discomfort referred to any sensation that would make them want to leave the setup \cite{fernandes2016combating}, including nausea, disorientation, and other symptoms captured by the SSQ. They answered the following question: “On a scale of 0 to 10, where 0 represents how you felt before starting and 10 means you want to stop, how do you feel now?” This approach, adapted from Fernandes and Feiner \cite{fernandes2016combating}, has been employed in previous VR and VST studies to facilitate real-time monitoring of discomfort and sickness throughout the trial \cite{adhanom2020effect,freiwald2018camera}. The discomfort score collected after the interaction task was considered as the ending discomfort score, and the average discomfort score was calculated using all the collected scores.

\subsubsection{Task Performance} In addition to the self-reported scores, we collected objective performance metrics for each task outlined in Section \ref{subsec:tasks}. For the typing task, we measured speed in characters per minute (CPM), calculated as the total number of correctly typed characters (including spaces) normalized to 60 seconds, and the error rate (ER). In the navigation task, we recorded navigation time in seconds and the number of cones dropped outside the designated drop zone (ER). For the interaction task, we tracked completion time and the number of correctly placed puzzle pieces. This allowed us to calculate the interaction performance metric of correctly placed puzzle pieces per minute (PPM).
\subsubsection{Qualitative Feedback} At the end of each trial, participants were asked to select their preferred VST condition and provide detailed feedback on their choice. Data from open-ended survey questions and observations were collected during and after participant experiences with the VST conditions. 

\subsection{Hypotheses}
Considering the previously described measures, we formulated the following hypotheses:\\
 \textbf{\hypertarget{hypo:H1}{H1}:} \Depthpassthrough passthrough (GAP) reduces cybersickness and subjective discomfort over \Directpassthrough passthrough (DP) \\
 \textbf{\hypertarget{hypo:H2}{H2}:} \Depthpassthrough passthrough (GAP) is preferred by users over \Directpassthrough passthrough (DP) \\
 \textbf{\hypertarget{hypo:H3}{H3}:} \Depthpassthrough passthrough (GAP) results in higher task performance compared to \Directpassthrough passthrough (DP) \\
 \textbf{\hypertarget{hypo:H4}{H4}:} DP and GAP induce a common set of symptoms which are not experienced under natural vision (NV)

\subsection{Procedure}
Before starting the trial, participants signed a consent form and completed a demographics questionnaire. Their IPD was measured with an optical digital pupilometer, and the HMD was adjusted to match their IPD. The experimenter then provided instructions and demonstrated the procedures for the typing, navigation, and interaction tasks. Participants were randomly assigned to one of six different condition orders, \edits{which stemmed from varying the sequence of NV, DP, and GAP,} determined using a balanced Latin square design. \edits{In each condition, participants completed the typing, navigation, and interaction tasks in a fixed order to focus on overall cybersickness rather than task-specific effects. Randomizing the task order could have introduced variability from headset wear time, conflating task-specific cybersickness with overall exposure effects.}

Participants wore the headset continuously until they finished all tasks for a given condition. Throughout each condition, the experimenter collected discomfort scores and noted down key observations. Participants filled out the SSQ before and after each condition. To allow cybersickness symptoms to subside, participants were required to take a break of at least 15 minutes between conditions. During these breaks, they had access to water and a space with windows.

After completing all conditions, participants answered open-ended survey questions to provide insights into their experiences. The interview began with a discussion of preferences between the VST conditions, followed by probing reasons for those preferences. The entire session, including the two 15-minute breaks, took approximately 90 minutes to complete.

% \subsection{Data Analysis} Descriptive statistics were used to potentially accentuate the differences in experienced cybersickness between conditions. Two researchers were tasked to analyse the qualitative data using a reflexive thematic analysis approach to uncover subtle and layered experiences and perceptions \cite{braun2006using,braun2019reflecting}. The choice of this methodology was motivated by previous studies that focused on understanding user experiences with VR technology \cite{chen2024d,tan2022understanding,knibbe2018dream}. The coders followed the six steps suggested by Braun and Clarke \cite{braun2006using}. The analysis followed an extensive review of the dataset to gain a deep understanding of the data and its context. Inter-rater reliability was computed using Cohen’s kappa \cite{cohen1960coefficient}. After this, the coders conducted a reflection session to identify insightful quotes and create a final classification on which they both agreed. This step allowed us to ensure that individual perspectives and biases did not greatly affect the analysis.

%% file: Sections/Results.tex
\section{Results}

\subsection{Quantitative Measures}

\subsubsection{Cybersickness} 
We report four sub-scores: Total Severity, Oculomotor, Nausea, and Disorientation calculated from the SSQ responses using the methodology established by Kennedy et al. \cite{kennedy1993simulator}. For our analysis, we used the differences between post-condition and pre-condition measurements as the dependent variable. The results of statistical analysis are summarized in Table \ref{tab:ssq}.

\begin{figure*}[!h]
  \centering
  \includegraphics[width=\textwidth]{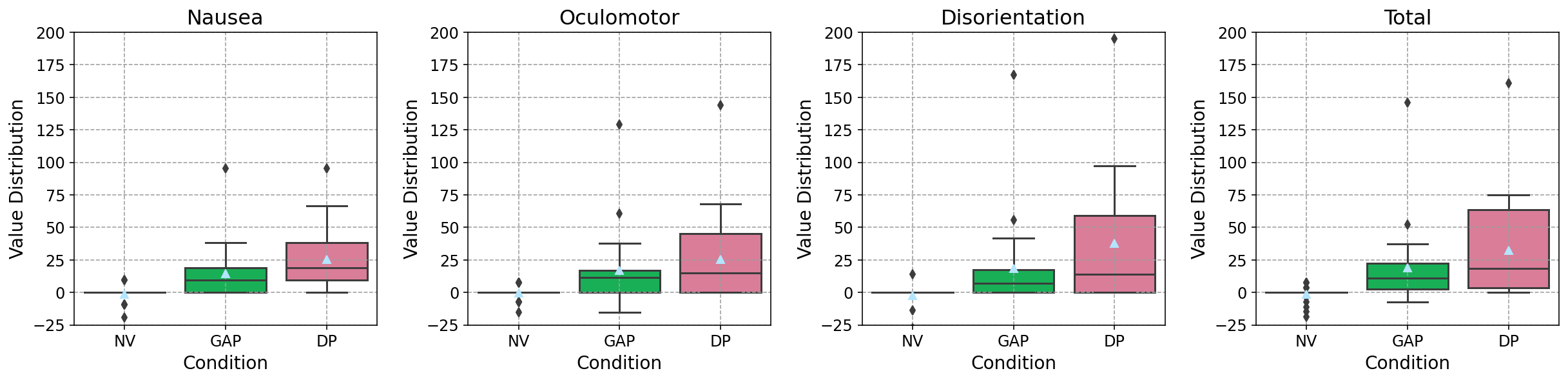}
  \caption{\textbf{SSQ} Box plots of Nausea, Oculomotor, Disorientation, and Total subscores of simulator sickness comparing all conditions. NV led to significantly lower N, O, D, and total sickness scores than both \DP and \GAP (p < 0.05). \GAP led to significantly lower N (p = 0.016), D (p = 0.029), and total (p = 0.011) sickness scores than \DP.}
  \Description{Box plots of the SSQ scores: Compared to NV and GAP, DP has a broader range and higher values across all scores: nausea, oculomotor, disorientation, and total.}
  \label{fig:ssq}
\end{figure*}

The Shapiro-Wilk test indicated that the SSQ scores were not normally distributed ($p$ < 0.001). Consequently, Friedman test was conducted for each SSQ sub-scores, which showed significant differences between conditions for Nausea ($\chi^2$ = 31.89, $p$ < 0.001), Oculomotor ($\chi^2$ = 25.97, $p$ < 0.001), Disorientation ($\chi^2$ = 23.89, $p$ < 0.001), and Total Severity ($\chi^2$ = 33.82, $p$ < 0.001). To identify specific condition differences, we performed pairwise Wilcoxon signed-rank tests with Holm-Bonferroni adjustment. NV resulted in significantly lower Nausea, Oculomotor, Disorientation, and Total Severity scores compared to both DP and GAP ($p$ < 0.05). \GAP led to significantly lower Nausea ($p$ = 0.016), Disorientation ($p$ = 0.029), and Total Severity ($p$ = 0.011) scores than \DP. Although \GAP had lower mean Oculomotor scores compared to \DP, the difference was not statistically significant.
Figure \ref{fig:ssq} displays the distribution of simulator sickness scores for all three conditions across sub-scores. Table \ref{tab:ssq} summarizes the \edits{mean, standard deviation, median, and interquartile range} of these sub-scores. We observed that disorientation disturbances contributed to simulator sickness the most, followed by oculomotor and nausea.

Inspired by Vovk et al. \cite{vovk2018simulator}, we examined individual SSQ symptoms in isolation to obtain a fine-grained understanding of VST induced cybersickness. As shown in Table \ref{tab:freq}, we observed that eyestrain and sweating were a major factor contributing to simulator sickness in VST in comparison to NV. The most reported symptoms for both \DP and \GAP were sweating, eyestrain, general discomfort, and headache. Wilcoxon signed-rank tests indicated that symptoms of general discomfort, sweating, eyestrain, headache, and blurred vision were significantly lower ($p<0.05$) for NV compared to both DP and GAP.

\begin{table*}[!h]
  \caption{\edits{Mean, standard deviation, median, and interquartile range} of simulator sickness for all conditions}
  \label{tab:ssq}
  \resizebox{\textwidth}{!}{
  \begin{tabular}{r|cccc|cccc|cccc|cccc}
    \toprule
    &\multicolumn{4}{c}{Nausea (N)}&\multicolumn{4}{c}{Oculomotor (O)}& \multicolumn{4}{c}{Disorientation (D)}& \multicolumn{4}{c}{Total Score}\\
    \midrule
     \edits{Condition} &Mean &SD &Median &IQR &Mean &SD &Median &IQR &Mean &SD &Median &IQR &Mean &SD &Median &IQR\\
    \midrule
    \textit{\edits{NV}} & \textit{-1.19} &\textit{5.72}
    &\edits{\textit{0}}
    &\edits{\textit{0}}
    &\textit{-0.63} &\textit{4.85}
    &\edits{\textit{0}}
    &\edits{\textit{0}}
    &\textit{-2.32} &\textit{8.68}
    &\edits{\textit{0}}
    &\edits{\textit{0}}
    &\textit{-1.40} &\textit{5.99}
    &\edits{\textit{0}}
    &\edits{\textit{0}}\\
    \DP &25.44 &22.82 &\edits{19.08} &\edits{28.62} &25.27 &33.09 &\edits{15.16} &\edits{45.48} &37.70 &47.96 &\edits{13.92} &\edits{59.16} &32.57 &36.79 &\edits{18.70} &\edits{59.84}\\
    \GAP &\textbf{14.71} &20.05 &\edits{\textbf{9.54}} &\edits{19.08} &\textbf{17.06} &27.79 &\edits{\textbf{11.37}} &\edits{17.05} &\textbf{18.56} &34.72 &\edits{\textbf{6.96}} &\edits{17.40} &\textbf{19.17} &29.97 &\edits{\textbf{11.22}} &\edits{19.64}\\
  \bottomrule
\end{tabular}
}
\end{table*}

\begin{table*}[!h]
  \caption{Mean, standard deviation, and percentage of participants (\%) with SSQ symptoms for all conditions. (N – Nausea, O – Oculomotor, D – Disorientation) }
  \label{tab:freq}
  \resizebox{\textwidth}{!}{%
  \begin{tabular}{r|ccc|ccc|ccc|ccc|ccc}
    \toprule
                         & \multicolumn{3}{c|}{NV}         & \multicolumn{3}{c|}{DP} & \multicolumn{3}{c|}{GAP} & \multicolumn{3}{c|}{DP $\cap$ GAP} &   &   &   \\ \hline
SSQ                      & Mean  & SD   & \% & Mean    & SD      & \%    & Mean     & SD      & \%    & Mean     & SD      & \%    & N & O & D \\ \hline
General discomfort       & -0.04 & 0.36 & 4\%             & 0.67    & 0.80    & 50\%               & 0.50     & 0.82    & 42\%               & 0.58     & 0.81    & 25\%               & x & x &   \\
Fatigue                  & 0.04  & 0.36 & 8\%             & 0.38    & 0.70    & 29\%               & 0.21     & 0.71    & 17\%               & 0.29     & 0.71    & 17\%               &   & x &   \\
Headache                 & -0.04 & 0.20 & 0\%             & 0.50    & 0.82    & 33\%               & 0.25     & 0.43    & 25\%               & 0.38     & 0.67    & 25\%               &   & x &   \\
Eyestrain                & 0.00  & 0.29 & 4\%             & 0.50    & 0.71    & 42\%               & 0.54     & 0.82    & 46\%               & 0.52     & 0.76    & 33\%               &   & x &   \\
Difficulty focusing      & 0.00  & 0.00 & 0\%             & 0.42    & 0.70    & 29\%               & 0.25     & 0.52    & 21\%               & 0.33     & 0.62    & 17\%               &   & x & x \\
Increased Salivation     & 0.00  & 0.00 & 0\%             & 0.04    & 0.20    & 4\%                & 0.04     & 0.20    & 4\%                & 0.04     & 0.20    & 0\%                & x &   &   \\
Sweating                 & 0.08  & 0.28 & 8\%             & 0.92    & 0.76    & 71\%               & 0.58     & 0.64    & 50\%               & 0.75     & 0.72    & 50\%               & x &   &   \\
Nausea                   & -0.08 & 0.28 & 0\%             & 0.50    & 0.65    & 42\%               & 0.25     & 0.53    & 21\%               & 0.38     & 0.60    & 17\%               & x &   & x \\
Difficulty concentrating & 0.00  & 0.00 & 0\%             & 0.33    & 0.55    & 29\%               & 0.17     & 0.47    & 13\%               & 0.25     & 0.52    & 13\%               & x & x &   \\
Fullness of the head     & 0.00  & 0.29 & 4\%             & 0.33    & 0.62    & 25\%               & 0.25     & 0.66    & 18\%               & 0.29     & 0.64    & 17\%               &   &   & x \\
Blurred vision           & -0.04 & 0.20 & 0\%             & 0.54    & 0.82    & 38\%               & 0.33     & 0.69    & 25\%               & 0.44     & 0.76    & 21\%               &   & x & x \\
Dizziness (eyes open)    & -0.04 & 0.20 & 0\%             & 0.50    & 0.71    & 42\%               & 0.08     & 0.28    & 8\%                & 0.29     & 0.58    & 8\%                &   &   & x \\
Dizziness (eyes closed)  & 0.00  & 0.00 & 0\%             & 0.30    & 0.61    & 21\%               & 0.04     & 0.35    & 8\%                & 0.17     & 0.51    & 8\%                &   &   & x \\
Vertigo                  & 0.00  & 0.00 & 0\%             & 0.13    & 0.33    & 13\%               & 0.13     & 0.33    & 13\%               & 0.13     & 0.33    & 8\%                &   &   & x \\
Stomach awareness        & -0.08 & 0.28 & 0\%             & 0.13    & 0.44    & 17\%               & 0.00     & 0.50    & 4\%                & 0.063    & 0.47    & 4\%                & x &   &   \\
Burping                  & 0.00  & 0.00 & 0\%             & 0.08    & 0.28    & 8\%                & 0.00     & 0.00    & 0\%                & 0.04     & 0.20    & 0\%                & x &   &   \\ 
  \bottomrule
\end{tabular}%
}
\end{table*}

\begin{figure*}[!ht]
  \centering
  \includegraphics[width=\textwidth]{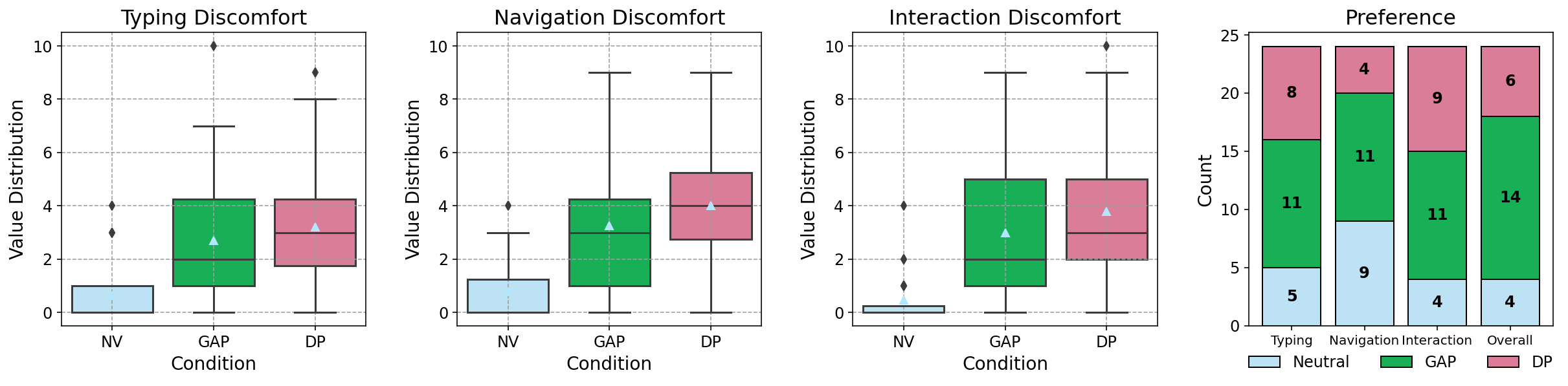}
  \caption{\textbf{Subjective Discomfort} Box plots of discomfort scores and preference for all conditions across the typing, navigation, and interaction tasks.  NV led to significantly lower discomfort scores than both \DP and \GAP (p < 0.05) across all tasks.  Compared to \DP, \GAP led to significantly lower discomfort scores across all tasks, i.e., typing (p = 0.046), navigation(p = 0.041), interaction (p = 0.022).}
  \label{fig:disc-all}
  \Description{Box plots of the subjective discomfort scores across the typing, navigation, and interaction tasks: Compared to NV and GAP, DP has higher values. However, GAP has a broader range. A stacked bar plot of user preference shows that GAP ranked first for all tasks and the overall experience.}
\end{figure*}

Subjective discomfort scores across all tasks and conditions were also tested for normality using the Shapiro-Wilk test, which showed non-normal distribution ($p$ < 0.001). Friedman tests revealed significant differences in discomfort scores between conditions for typing ($\chi^2$ = 26.60, $p$ < 0.001), navigation ($\chi^2$ = 28.00, $p$ < 0.001), interaction ($\chi^2$ = 33.34, $p$ < 0.001), and average discomfort ($\chi^2$ = 33.23, $p$ < 0.001). Pairwise Wilcoxon signed-rank tests with Holm-Bonferroni adjustment demonstrated that NV led to significantly lower discomfort scores across all tasks compared to both \DP and \GAP ($p$ < 0.05). NV also resulted in significantly lower average discomfort scores compared to both \DP and \GAP ($p$ < 0.05). GAP led to significantly lower discomfort scores across all tasks compared to DP, including typing ($p$ = 0.046), navigation ($p$ = 0.041), and interaction ($p$ = 0.022), as well as significantly lower average discomfort scores ($p$=0.016). Figure \ref{fig:disc-all} illustrates the distribution of discomfort scores for all three conditions across tasks, with navigation showing the highest mean discomfort, followed by interaction. Means and standard deviations of the discomfort scores are summarized in Table \ref{tab:subj-disc}.

\begin{table}[!ht]
  \caption{Mean and standard deviation of subjective discomfort across tasks for all conditions}
  \label{tab:subj-disc}
  \resizebox{\columnwidth}{!}{
  \begin{tabular}{r|cc|cc|cc|cc}
    \toprule
    &\multicolumn{2}{c}{Typing}&\multicolumn{2}{c}{Navigation}& \multicolumn{2}{c}{Interaction}&\multicolumn{2}{c}{Average}\\
    \midrule
     \edits{Condition} &Mean &SD &Mean &SD &Mean &SD &Mean &SD\\
    \midrule
    \textit{\edits{NV}} & \textit{0.63} &\textit{1.11} &\textit{0.79} &\textit{1.12} &\textit{0.46} &\textit{0.96} &\textit{0.63}	&\textit{0.99}\\
    \DP &3.21 &2.47 &4.00 &2.57 &3.79 &2.63 &3.67 &2.38\\
    \GAP &\textbf{2.71} &2.57 &\textbf{3.25} &2.73 &\textbf{3.00} &2.65 &\textbf{2.99} &2.52\\
  \bottomrule
\end{tabular}
}
\end{table}

% \begin{figure}[h]
%   \centering
%   \includegraphics[width=\linewidth]{images/subj-disc.png}
%   \caption{Subjective discomfort scores comparing all conditions across the typing, navigation, and interaction tasks.}
%   \label{fig:subj-disc}
%   \Description{A woman and a girl in white dresses sit in an open car.}
% \end{figure}

% \begin{table}[!ht]
%   \caption{Mean and standard deviation of simulator sickness for all conditions (Post)}
%   \label{tab:freq}
%   \begin{tabular}{r|cc|cc|cc|cc}
%     \toprule
%     &\multicolumn{2}{c}{Nausea (N)}&\multicolumn{2}{c}{Oculomotor (O)}& \multicolumn{2}{c}{Disorientation (D)}& \multicolumn{2}{c}{Total Score}\\
%     \midrule
%      &Mean &SD &Mean &SD &Mean &SD &Mean &SD\\
%     \midrule
%     \textit{Natural vision} & \textit{3.18} &\textit{5.95} &\textit{4.42} &\textit{5.75} &\textit{0.58} &\textit{2.78} &\textit{3.58} &\textit{4.25}\\
%     Planar projection VST &28.22 &24.24 &28.74 &32.23 &38.28 &47.76 &35.53 &36.61\\
%     Depth reprojection VST &\textbf{20.27} &22.25 &\textbf{24.32} &30.55 &\textbf{24.94} &43.46 &\textbf{26.65} &34.42\\
%   \bottomrule
% \end{tabular}
% \end{table}

\begin{table*}[!ht]
  \caption{The results of statistical analysis for the SSQ and subjective discomfort scores ($^{\ast} p < 0.001$; $^{\ast\ast} p < 0.05$)}
  \label{tab:stats}
  \resizebox{0.7\textwidth}{!}{
  \begin{tabular}{ll|l|l}
    \toprule
     \multicolumn{2}{l|}{\textbf{Dependent Variables}} & \textbf{Friedman} & \textbf{Wilcoxon signed-rank} \\
    \midrule
    \textbf{SSQ} & Nausea & $\chi^2 = 31.89^{\ast}$ & \DP > NV$^{\ast\ast}$, \GAP > NV$^{\ast\ast}$, \DP > \GAP$^{\ast\ast}$ \\
    & Oculomotor & $\chi^2 = 25.97^{\ast}$ & \DP > NV$^{\ast\ast}$, \GAP > NV$^{\ast\ast}$ \\
    & Disorientation & $\chi^2 = 23.89^{\ast}$ & \DP > NV$^{\ast\ast}$, \GAP > NV$^{\ast\ast}$, \DP > \GAP$^{\ast\ast}$ \\
    & Total & $\chi^2 = 33.82^{\ast}$ & \DP > NV$^{\ast\ast}$, \GAP > NV$^{\ast\ast}$, \DP > \GAP$^{\ast\ast}$ \\
    \midrule
    \textbf{Discomfort} & Typing & $\chi^2 = 26.60^{\ast}$ & \DP > NV$^{\ast\ast}$, \GAP > NV$^{\ast\ast}$, \DP > \GAP$^{\ast\ast}$ \\
    & Navigation & $\chi^2 = 28.00^{\ast}$ & \DP > NV$^{\ast\ast}$, \GAP > NV$^{\ast\ast}$, \DP > \GAP$^{\ast\ast}$ \\
    & Interaction & $\chi^2 = 33.34^{\ast}$ & \DP > NV$^{\ast\ast}$, \GAP > NV$^{\ast\ast}$, \DP > \GAP$^{\ast\ast}$ \\
    & Average & $\chi^2 = 33.23^{\ast}$ & \DP > NV$^{\ast\ast}$, \GAP > NV$^{\ast\ast}$, \DP > \GAP$^{\ast\ast}$ \\
    \bottomrule
  \end{tabular}
  }
\end{table*}

\subsubsection{Task Performance} Shapiro-Wilk tests indicated that the data for CPM, typing error rate, and navigation error rate were not normally distributed ($p$ < 0.001). Friedman tests revealed significant differences in CPM ($\chi^2$ = 20.33, $p$ < 0.001) and typing error rate ($\chi^2$ = 6.10, $p$ < 0.001) between conditions. However, no significant differences were found in navigation error rate. Pairwise Wilcoxon tests with Holm-Bonferroni adjustment showed that both DP and GAP had significantly decreased CPM compared to NV ($p$ < 0.001), but no significant difference was found between DP and GAP. Typing error rate did not differ significantly between conditions.

Data for navigation time and PPM were normally distributed for NV, DP, and GAP ($p$ > 0.001). A repeated-measures ANOVA revealed significant differences in both navigation time (\( F(2, 46) = 27.373 \), \( p < 0.001 \), \( \eta^2 = 0.543 \)) and PPM (\( F(2, 46) = 8.88 \), \( p < 0.001 \), \(\eta^2 = 0.279 \)). Post-hoc paired t-tests with Holm-Bonferroni adjustment showed that navigation time and PPM were significantly different for NV compared to both DP and GAP (\( p < 0.05 \)). No significant differences were found between DP and GAP for navigation time (\( p = 0.779 \)) or PPM (\( p = 0.880 \)).

Although differences between DP and GAP were not statistically significant, GAP demonstrated improved mean typing CPM, typing error rate, navigation time, and navigation error rate compared to DP. Means and standard deviations of the task performance scores are summarized in Table \ref{tab:perf}.

\begin{figure*}[!ht]
  \centering
  \includegraphics[width=\textwidth]{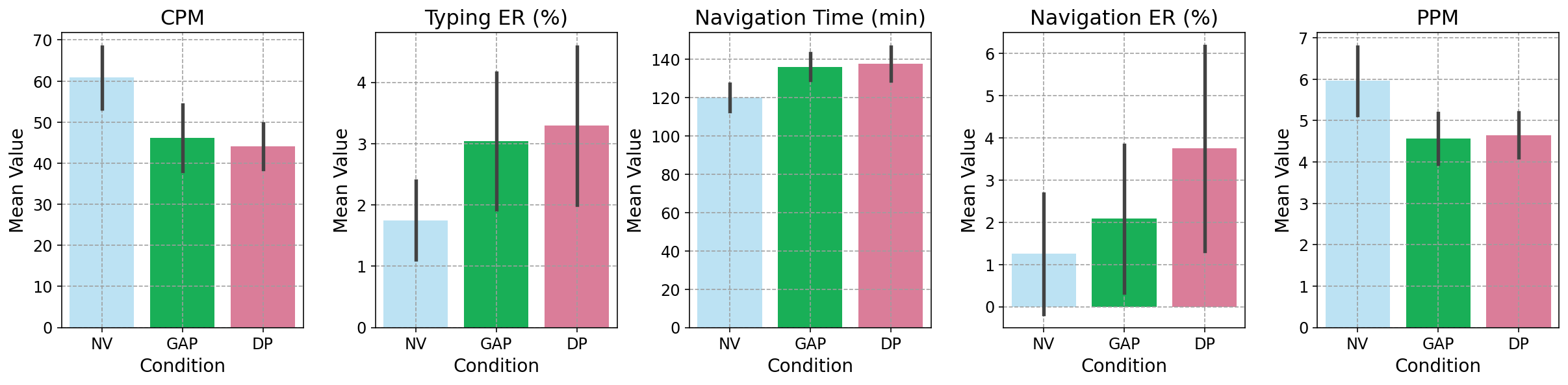}
  \label{fig:task-perf}
  \caption{\textbf{Task performance results:} (a) Typing speed (characters per minute), (b) Typing error rate (\%), (c) Navigation time (min), (c) Navigation error rate (\%), and (d) Puzzle speed (pieces per minute).}
  \Description{Bar plots of the task performance scores. The trends in the means are as follows. CPM: NV>GAP>DP, Typing ER (\%): NV<GAP<DP, Navigation Time (min): NV<GAP<DP, Navigation Error (\%): NV<GAP<DP, PPM: NV>DP>GAP. DP mean values are very slightly different than GAP mean values.}
\end{figure*}

\begin{table*}[!ht]
  \caption{Mean and standard deviation of task performance scores (CPM - characters per minute, ER - error rate in typing and navigation tasks; PPM - pieces per minute in interaction task) for all conditions. The best task performance is highlighted in bold.}
  \label{tab:perf}
  \resizebox{0.7\textwidth}{!}{
  \begin{tabular}{r|cc|cc|cc|cc|cc}
    \toprule
    &\multicolumn{4}{c}{Typing}|&\multicolumn{4}{c}{Navigation}|& \multicolumn{2}{c}{Interaction}\\
    \midrule
    &\multicolumn{2}{c}{CPM}&\multicolumn{2}{c}{ER (\%)}|& \multicolumn{2}{c}{Time (s)}&\multicolumn{2}{c}{ER (\%)}|&\multicolumn{2}{c}{PPM}\\
    \midrule
     \edits{Condition} &Mean &SD &Mean &SD &Mean &SD &Mean &SD &Mean &SD\\
    \midrule
    \textit{\edits{NV}} & \textit{60.83} &\textit{17.92} &\textit{1.75} &\textit{1.51} &\textit{120.13} &\textit{16.67} &\textit{1.25}	&\textit{3.38} &5.97 &1.95\\
    \DP &44.13 &13.04 &3.29 &3.06 &137.76 &21.22 &3.75 &5.76 &\textbf{4.65} &1.27\\
    \GAP &\textbf{46.21} &19.20 &\textbf{3.04} &2.65 &\textbf{136.21} &16.50 &\textbf{2.08} &4.15 &4.57 &1.44\\
  \bottomrule
\end{tabular}
}
\end{table*}

\subsection{Qualitative Feedback}
\label{subsec:qualfeedback}
% In this section, we present the results corresponding to the qualitative data collected using the open-ended survey. We first present our findings related to participant preference. We then discuss our thematic analysis to capture participants' experiences with the two VST conditions. The analysis provides detailed insights into discomfort, cybersickness, and perceived visual artifacts in each condition. We explore how the two VST modes compare in terms of their impact on user experience. 
% Table \ref{tab:thematic} displays the five identified themes and codes as well as the number of related text segments.  The kappa value $\kappa$ was 0.92, which shows an almost perfect level of agreement.

\subsubsection{\textbf{Users Prefer GAP}} Based on overall experience, participants generally preferred \GAP over \DP as show in Figure \ref{fig:disc-all}. For typing, 46\% participants preferred \GAP while 33\% preferred DP. Similarly, participants preferred \GAP for both the tasks navigation (GAP: 46\% vs. DP: 17\%) and interaction (GAP: 46\% vs. DP: 38\%).

\subsubsection{\textbf{GAP Reduces Cybersickness and Discomfort Compared to DP}}
User feedback indicated that DP led to more frequent cybersickness symptoms and discomfort , such as dizziness, eyestrain, and headache compared to GAP. This is consistent with the quantitative SSQ results, which showed significantly higher total cybersickness scores for DP. For instance, P4 commented, \textit{"The passthrough with [DP] was more straining on my eyes,"} and noted, \textit{"I would get tired of using this mode over a long period."}
% User feedback indicated that DP led to more frequent cybersickness symptoms and discomfort (27/38), such as dizziness (8/11), eyestrain (4/5), and headache (3/3) compared to GAP. This is consistent with the quantitative SSQ results, which showed significantly higher total cybersickness scores for DP. For instance, P4 commented, \textit{"The passthrough with [DP] was more straining on my eyes,"} and noted, \textit{"I would get tired of using this mode over a long period."}

% Although some symptoms were reported in open-ended survey questions, additional symptoms emerged from the SSQ (Table \ref{tab:freq}), suggesting that the SSQ captures a more comprehensive range of symptoms.  

\subsubsection{\textbf{Sensory Conflict in DP Causes Discomfort}}
Several participants reported a mismatch between vision and motion with DP, aligning with the sensory-conflict theory, the most accepted explanation for motion sickness. This was brought up more frequently in DP than in GAP by the participants, supporting our hypothesis that DP exaggerates motion effects. P17 described their experience: \textit{"What I see does not match what is happening...Do you trust what you see or how you move? These don't overlap, and I don't know which one to trust."} P12 also noted, \textit{"[DP] is worse when there is head motion, especially when completing the puzzle, because everything feels like it is moving."} 

\subsubsection{\textbf{Geometry Enhances Spatial Awareness}}
Participants reported impaired spatial awareness and depth perception with DP compared to GAP. Observations from the researcher confirmed that participants had to move closer to cones during the navigation task with DP due to inaccurate depth cues. User feedback supported these observations. For example, P17 said, \textit{"Depth perception was worse with [DP], and I tested that with the cone experiment. With [GAP], I could toss the cone in and not miss but with [DP] I had to be more careful about where to put the cone."} Additionally, unstable gait and more frequent collisions with furniture were noted in DP: 
\textit{"I almost knocked things over. I had to be more cognizant of my steps around the furniture and not hurt myself with the furniture. With [GAP], I did not have that problem."}

\subsubsection{\textbf{Artifacts in VST}}
Artifacts such as motion blur with DP and warping with GAP were recurrent themes. P17 described the experience with DP: 
\textit{"It is similar to car sickness because what I see does not match what is happening...The mismatch between vision and motion is more noticeable with [DP] because of the blur. It is another layer of having your brain decode the blurry image to understand the lock between perception and motion"}. While most participants preferred GAP for all tasks, some favored DP for typing due to warping issues on the keyboard: \textit{"I prefer \GAP for movements but \DP for typing because of the distortions on the keys"} (P1). A few participants adapted to warping over time:\textit{"I noticed distortions before starting to walk but it became better during the experience"} (P5). Warping was particularly noticeable at the edges, with feedback such as,
\textit{"[GAP] is less dizzying but feels like swimming in water. It feels good as long as I am looking in the center"} (P12). \edits{Furthermore, we also asked participants to report the reasoning for preferring a passthrough mode (\DP vs \GAP), wherein out of 25 participants recruited for our study, only 18 provided reasons for disliking a particular mode. 10 of these participants specifically indicated that image distortions (warping on objects and edges) were bothersome. This finding further emphasizes on the importance of measuring and minimizing warping artifacts for designing better passthrough systems.}

\subsubsection{\textbf{Avenues for Future Research}}
Participant feedback also helped identify several areas for future research into enhancing comfort in VST. Four issues emerged: frame drops, overexposed images, latency, and blurry vision.  Several participants reported that slight delays when moving their heads caused nausea and discomfort. For instance, P1 noted, \textit{"The slight lagging when moving my head causes nausea"}. Blurry vision was frequently cited as a significant issue, particularly with DP. Out of 93 comments, 18 specifically mentioned blurry vision, with a higher incidence observed in DP. Participants described how blurry visuals affected their ability to perform tasks comfortably. P19 observed, 
\textit{"Overall, my vision was less clear with [DP], especially with the typing challenge...I had to squint longer and go closer to the screen to be able to see the words."} 

%% file: Sections/Discussion.tex
\section{Discussion}
Our work primarily focuses on the impact of reprojection in passthrough on cybersickness and discomfort when using a VST HMD. Below, we first discuss the importance and impact of the proposed metrics in designing passthrough systems. We then discuss the main results of our user study and provide implications for improving VST systems. Lastly, we provide directions for future work while discussing the limitations of our work.

\subsection{Benchmarking and Technical Metrics}

One of our main findings is the improvement in cybersickness experienced by the users with GAP over DP. However, it is very expensive and time consuming to run user studies for evaluating comfort for every aspect of reprojection such as depth estimation network, device calibration, and with different tasks. Moreover, we expect small changes in GAP to only have small changes in the SSQ measurements. In this light, we proposed two metrics which directly aim to evaluate the two fundamental aspects of passthrough i.e. (i) perceived location of objects on the HMD display and (ii) the warping artifacts like stretching and bending of objects. We utilize these metrics to select the right geometry estimation method. 
Our spatial reprojection error metric both evaluates and elucidates perceived geometrical errors. Further, the warping error metric quantifies and thus enables minimization of added distortions that could alter the shape of rigid objects in the reprojected imagery. Together, these metrics can fundamentally enhance GAP systems, with user studies employed judiciously.
% We note that our spatial reprojection error metric not only evaluates but explains perceived geometrical errors. 
% In addition, the warping error metric allows quantifying and therefore minimizing added distortions that could change shape of rigid objects in the reprojected imagery. 
% In conjunction, these two metrics can therefore be utilized to fundamentally improve \depthpassthrough passthrough systems and relying on user studies judiciously.

\subsection{GAP Mitigates Cybersickness}
The results of our user study showed a signifcant reduction in  nausea, disorientation, and total scores of simulator sickness with GAP compared to DP. GAP also led to significantly lower subjective discomfort scores across all tasks: typing, navigation, and interaction as well as a significantly lower average discomfort score. These two findings confirm (\hyperlink{hypo:H1}{H1}). We identified that \GAP exposed participants to low levels of cybersickness while \DP exposed participants to moderate levels of cybersickness \cite{hale2014handbook,stanney2020identifying}. 
Participants who experienced greater discomfort with DP showed a clear preference for GAP, which was ranked first in all tasks and for the overall experience, leading to
acceptance of (\hyperlink{hypo:H2}{H2}). Qualitative feedback provided additional insights into the sources of discomfort associated with DP. Participants experienced varying levels of cybersickness depending on the task. Specifically, many preferred GAP for navigation tasks due to a noticeable mismatch between visual and inertial cues, which was evident in their unstable gait and impaired spatial awareness. In contrast, some participants were more affected by head motion during the interaction task as they assembled the puzzle. Despite these subjective preferences, no significant differences were found in the five objective measures of task performance, leading to the rejection of (\hyperlink{hypo:H3}{H3}).
% We considered the differences between the post-condition and pre-condition measurements as the dependent variable. 
\subsection{Distinctive Features of Cybersickness in VST}  
As expected, natural vision outperformed both VST conditions in cybersickness and discomfort as well as task performance. In line with previous work on VST \cite{de2024visual}, our results revealed a VST symptom profile for cybersickness that is distinguished from other types of motion sickness. Based on the results of the SSQ, cybersickness in VST showed a  Disorientation > Oculomotor > Nausea profile. In contrast, previous studies have reported different symptom profiles for other types of motion sickness: VR cybersickness typically follows a Disorientation > Nausea > Oculomotor pattern, simulator sickness follows an Oculomotor > Nausea > Disorientation pattern, and sea sickness follows a Nausea > Oculomotor > Disorientation pattern \cite{de2024visual,gallagher2018cybersickness,somrak2019estimating}. Symptoms of general discomfort, sweating, eye strain, headache, and blurred vision were significantly lower for NV compared to both DP and GAP. This finding supports (\hyperlink{hypo:H4}{H4}). 
While symptoms of nausea, difficulty concentrating, difficulty focusing, and dizziness (eyes open) are significantly lower with NV compared to DP, they show no significant differences between NV and GAP. 
This suggests that GAP effectively reduces some aspects of cybersickness but does not fully bridge the gap to natural vision. 
The most frequent symptoms experienced with both VST conditions are sweating followed by eyestrain, headache, and general discomfort. 
Interestingly, these symptoms are also the most frequent symptoms identified by Vovk et al. in their simulator sickness evaluation with the Microsoft HoloLens, an AR OST HMD \cite{vovk2018simulator}. 
The least frequent ones are burping and increased salivation.
\subsection{Implications for VST HMD Design}
Our study demonstrates that GAP significantly enhances user comfort compared to DP. This implies that VST HMD design should incorporate GAP to improve user comfort and facilitate the adoption of VST technology. However, we must be aware that GAP introduces additional computational demands compared to DP. Future design efforts should strive for a balance that maintains the benefits of GAP while minimizing the impact on system performance and latency. 
% For instance, efficient machine learning models for estimating depth or geometry can be utilized to further reduce the computation needs for GAP based systems. 
Lastly, we perform depth estimation to facilitate the reprojection process, but many recent view synthesis approaches \cite{kerbl3Dgaussians, mildenhall2020nerf, li2023dynibar} have come up which can be optimized for real-time on-device usage to devise geometrically accurate and comfortable passthrough solutions.
% \edits{\subsection{Impact of Reprojection Error on Comfort}We also conducted a small pilot experiment with 6 new participants demonstrating the impact of our metrics on comfort. We considered two passthrough systems with varying depth quality (GAP - HighQ, GAP - LowQ) and ran these on our protocol. GAP - HighQ is same as GAP we employed for our user study in previous sections, and GAP - LowQ uses a degraded version of the depth than GAP - HighQ such that it has about 8 times higher reprojection error and 28\% smaller warping error. We observed that our protocol could clearly distinguish improved comfort between GAP - HighQ and GAP - LowQ (sickness scores increase by a factor of 2 on average with degradation of depth). This indicates that increased spatial reprojection error has raised user discomfort even with lesser warping artifacts. While precisely calculating correlation or causality is out of scope for our work, we hope that similar studies with larger sample sizes can become great avenues for future research.}
\subsection{Limitations and Future Work}
Apart from reprojection and geometry perception, there are a wide range of other factors in VST HMDs which affect the user comfort such as the latency and temporal consistency of scene updates, field of view, physical fit of the device, image quality factors such as noise and sharpness levels, accuracy of display colors, camera optics of the device among others. 
This is reflected in the qualitative feedback where latency, frame drops, overexposed images, and blurry vision are recurrently mentioned. Particularly, blurry vision is the most frequently cited issue with a higher incidence in DP. Essentially, all the discrepancies between human natural vision and VST HMD need to be minimized for a comfortable VST experience. %Modern camera and imaging pipelines are typically created for digital photography use case but not designed from first principles for VST HMD comfort. 
% Perceived scene brightness and dynamic range are dictated by a combination of VST camera’s hardware and software image processing pipelines as well as the choice of digital display technology such as OLED vs LCD. Either too high or too low perceived brightness outside of an optimal range may cause eye strain in prolonged usage sessions [REF]. What may work well for purely virtual content in a VR HMD device may not work as well for VST in an AR or XR device. Accuracy of display colors, image noise, and sharpness also affects user comfort. Most software image processing pipelines are created and tuned for photography use cases for modern smartphones and digital cameras. For HMD VST, low latency and perceptual temporal consistency are far more important. Designing camera hardware and software image processing specifically for the HMD VST use case and treating user comfort as first principle has received much less attention in the literature [REF].\newline
This suggests the need for more research to understand the impact of these factors on mitigating user discomfort and cybersickness. 

\edits{Our findings emphasize that motion sickness mitigation should not be taken for granted in VST systems. While GAP significantly reduces motion sickness compared to DP, it does so at a computational cost, requiring additional power and pipeline complexity. However, the value that GAP adds to user comfort and sickness mitigation justifies this cost. We encourage future work to explore the balance between system performance, energy efficiency, and user comfort in XR systems. GAP pipelines that dynamically scale computational demands based on user motion or environmental factors could further optimize this balance, reducing unnecessary energy expenditure while maintaining the benefits of sickness mitigation.}

\edits{One limitation of our study was the within-subjects design with shorter washout periods between conditions. We acknowledge that assuming participants can fully recover from cybersickness symptoms within the same session may be risky as recovery can take up to 24 hours for some individuals. However, we adopted this approach to directly capture user preferences across conditions and to balance the practical constraints of participant recruitment and retention. To mitigate carryover effects, we implemented proper counterbalancing and provided participants with a 15-minute break between conditions, informing them they could extend this break if needed. None of the participants reported requiring additional time. Furthermore, we analyzed the differences between post and pre task SSQ scores for each condition to isolate the relative cybersickness impact.}

\edits{Assessing passthrough systems is challenging due to their complexity and the interdependency between various factors (framerate, power consumption, depth quality, rendering pipeline, Motion2Photon latency), making it hard and potentially redundant to evaluate or correlate a single component’s impact on comfort.} While in this work, we solely focused on isolating the impact of GAP, we hope that our proposed protocol would encourage future work in understanding the impact of other factors on visually-induced sickness in VST HMDs. \edits{Given the growing prevalence of VST technology, we believe it is crucial to further investigate comfort and usability, which remain underexplored.} Finally, we expect future work to correlate our proposed metrics to the user comfort. This could bring new insights and potentially new measures for assessing passthrough comfort.

% Lastly, our study did not include overlaid virtual information as we
% aimed at focusing on the actual VST methods. Future work can look at cybersickness induced by virtual content in VST such as world-locked content.

%% file: Sections/Conclusion.tex
\section{Conclusion}
In this work, we investigated the impact of reprojection algorithms on user discomfort and cybersickness in VST HMDs. We first proposed metrics to evaluate geometry correctness and warping artifacts generated by \depthpassthrough reprojection. We show how these metrics allow comparison between different passthrough algorithms with a focus on the percevied scene geometry and object shapes. 

We presented a comprehensive protocol aimed at evaluating visually-induced discomfort and cybersickness in VST HMDs through key use cases. Our results indicate that GAP significantly reduces nausea, disorientation, and total scores of cybersickness as well as subjective discomfort scores as compared to DP. Notably, our findings revealed specific symptoms unique to VST systems. Following sweating, eyestrain, a symptom in the oculomotor group, emerged as the main factor contributing to simulator sickness in VST. This suggests that cybersickness in VST systems is a complex phenomenon, distinct from both VR cybersickness and traditional motion sickness. Although the levels of cybersickness observed were generally low to moderate, these symptoms can influence user acceptance and overall experience. The feedback collected from users revealed new avenues for further research and improvements in VST HMD comfort. We hope that our comprehensive protocol sets a foundation for future studies aimed at refining these systems and enhancing user comfort in VST technologies.